\documentclass[aps,prl,nofootinbib,twocolumn,showpacs,superscriptaddress,letterpaper,amsmath,amssymb]{revtex4}
\usepackage{graphicx}  
\usepackage{dcolumn}   
\usepackage{bm}        
\usepackage{amssymb}   
\usepackage{feynmf}
\usepackage{hyperref}
\usepackage{slashed}
\usepackage{multirow}
\usepackage{color}
\usepackage{array}

\newcolumntype{L}[1]{>{\raggedright\let\newline\\\arraybackslash\hspace{0pt}}m{#1}}
\newcolumntype{C}[1]{>{\centering\let\newline\\\arraybackslash\hspace{0pt}}m{#1}}
\newcolumntype{R}[1]{>{\raggedleft\let\newline\\\arraybackslash\hspace{0pt}}m{#1}}

\def\gsim{\lower0.5ex\hbox{$\:\buildrel >\over\sim\:$}}
\def\lsim{\lower0.5ex\hbox{$\:\buildrel <\over\sim\:$}}

\newcommand{\be}{\begin{equation}}
\newcommand{\ee}{\end{equation}}
\newcommand{\bea}{\begin{eqnarray}}
\newcommand{\eea}{\end{eqnarray}}

\newcommand{\nbox}{{\,\lower0.9pt\vbox{\hrule \hbox{\vrule height 0.2 cm
\hskip 0.2 cm \vrule height 0.2 cm}\hrule}\,}}

\def\sub#1{_{\lower.25ex\hbox{$\scriptstyle#1$}}}

\newskip\zatskip \zatskip=0pt plus0pt minus0pt
\def\matth{\mathsurround=0pt}
\def\lsim{\mathrel{\mathpalette\atversim<}}
\def\gsim{\mathrel{\mathpalette\atversim>}}
\def\sigv{\ifmmode \langle\sigma v\rangle\else $\langle\sigma v\rangle$\fi}
\newskip\zatskip \zatskip=0pt plus0pt minus0pt
\def\matth{\mathsurround=0pt}
\def\lsim{\mathrel{\mathpalette\atversim<}}
\def\gsim{\mathrel{\mathpalette\atversim>}}
\def\atversim#1#2{\lower0.7ex\vbox{\baselineskip\zatskip\lineskip\zatskip
  \lineskiplimit
  0pt\ialign{$\matth#1\hfil##\hfil$\crcr#2\crcr\sim\crcr}}}

\begin{document}

\thispagestyle{empty}
\vspace*{-3.5cm}

\vspace{0.5in}

\title{Jet Flavor Classification in High-Energy Physics with Deep Neural Networks}

\begin{center}
\begin{abstract}
  
Classification of jets as originating from light-flavor or heavy-flavor quarks is an important task for inferring the nature of particles produced in high-energy collisions.  The large and variable dimensionality of the data provided by the tracking detectors makes this task difficult. The current state-of-the-art tools require expert data-reduction to convert the data into a fixed low-dimensional form that can be effectively managed by shallow classifiers.  We study the application of deep networks to this task, attempting classification at several levels of data, starting from a raw list of tracks.  We find that the highest-level lowest-dimensionality expert information sacrifices information needed for classification, that the performance of current state-of-the-art taggers can be matched or slightly exceeded by deep-network-based taggers using only track and vertex information, that classification using only lowest-level highest-dimensionality tracking information remains a difficult task for deep networks, and that adding lower-level track and vertex information to the classifiers provides a significant boost in performance compared to the state-of-the-art.

\end{abstract}
\end{center}

\author{Daniel Guest\footnote{The first two authors contributed equally}}
\affiliation{Department of Physics and Astronomy, University of  California, Irvine, CA 92697}
\author{Julian Collado$^*$}
\affiliation{Department of Computer Science, University of  California, Irvine, CA 92697}
\author{Pierre Baldi}
\affiliation{Department of Computer Science, University of  California, Irvine, CA 92697}
\author{Shih-Chieh Hsu}
\affiliation{Department of Physics, University of Washington, Seattle, WA 98195}
\author{Gregor Urban}
\affiliation{Department of Computer Science, University of  California, Irvine, CA 92697}
\author{Daniel Whiteson}
\affiliation{Department of Physics and Astronomy, University of  California, Irvine, CA 92697}

\date{\today}

\pacs{}
\maketitle


\section{Introduction}

The search for new particles and interactions at the energy frontier is a rich program with enormous discovery potential. The power to discover this hypothetical new physics  relies crucially on the ability to infer the nature of the interaction and the particles produced from the data provided by the detectors which surround the point of collision. One critical element is jet flavor classification, the distinction between hadronic jets produced from light-flavor and heavy-flavor quarks. Such classification plays a central role in identifying heavy-flavor signals and reducing the enormous backgrounds from light-flavor processes~\cite{Aad:2016kww,Aad:2016shx}.

Jets originating from heavy-flavor quarks tend to produce longer-lived particles than those found in jets from light-flavor quarks; these long-lived particles have  decays which are displaced from the primary vertex. To identify such vertices, the central tracking chamber measures the trajectories of charged particles which allows for the reconstruction of vertex locations.   The large and varying number of particles in a jet leads to a difficult classification problem with large and variable dimensionality without a natural ordering. The first step in typical approaches involves vertex-finding algorithms~\cite{Waltenberger:2008zza}, which transform the task into one of reduced, but still variable, dimensionality. Finally, most state-of-the-art jet flavor classification tools used by experiments~\cite{Aad:2015ydr,Chatrchyan:2012jua} rely heavily on expert-designed features which fix and further reduce the dimensionality before applying shallow machine-learning techniques. Such techniques have excellent performance, but are primarily motivated by historical limitations in the ability of shallow learning methods to handle high- and variable-dimensionality datasets. 

 Recent applications of deep learning to similar problems in high-energy physics~\cite{baldi_searching_2014, baldi_enhanced_2015,sadowski_deep_2014,Baldi:2016fql}, combined with the lack of a clear analytical theory to provide dimensional reduction without loss of information,  suggests that deep learning techniques applied to the lower-level higher-dimensional data could yield improvements in the performance of jet-flavor classification algorithms.  General methods for designing and applying recurrent and recursive neural networks to problems with data of variable size or structure have been developed in Refs.~\cite{Baldi:1996:HMH:1362127.1362136,goller,Frasconi:1998:GFA:2325763.2326281,gers,Baldi:2003:PDL:945365.945379}, and applied systematically to a variety of problems ranging from natural language processing~\cite{socher}, to protein structure prediction~\cite{brunak,tegge,dilena,magnan} to prediction of molecular properties~\cite{lusci,NIPS2015_5954} and to the game of go~\cite{Wu20081392}; previous studies have discussed the extension of such strategies to tasks involving tracks in high energy physics~\cite{dslhc1,dslhc2}.

In this paper, we apply several deep learning techniques to this problem using a structured dataset with features at three levels of processing (tracks, vertices, expert), each of which is a strict function of the previous level(s).  The data at the highest level of processing, with smallest dimensionality, is intended to mirror the typical approach used currently by experimental collaborations.   The multi-layered structure of the dataset allows us to draw conclusions about the information loss at each stage of processing, and to gauge the ability of machine learning tools to find solutions in the lower- and higher-dimensional levels. These lessons can guide the design of flavor-tagging algorithms used by experiments.


\section{Classification and Dimensionality}

The task of the machine learning (ML) algorithm is to identify a function $f(\bar{x}): {\rm I\!R}^N \rightarrow {\rm I\!R}^1$ whose domain is the observed data at some level of processing (with potentially very large dimensionality $N$) and which evaluates to a single real value that contains the information necessary to perform the classification.   Perfect classification is not expected; instead, the upper bound is performance which matches classification provided by the true likelihood ratio between heavy-flavor ($b$) and light-flavor quarks ($q$): $P(\bar{x}|b)/P(\bar{x}|q)$ evaluated in the high-dimensional domain.

Though we lack knowledge of an analytical expression for the likelihood, in principle one could recover such a function from labeled datasets with trivial algorithms, by estimating the likelihood directly in the original high-dimensional space. In practice, this requires an enormous amount of data, making it impractical for problems with anything but the smallest dimensionality in their feature space. 

Machine learning  plays a critical role in approximating the function $f(\bar{x})$ which reduces the dimensionality of the space to unity by finding the critical information needed to perform the classification task. Such a function may disregard some of the information from the higher-dimensional space if it is not pertinent to the task at hand.  However, for very high dimensional spaces (greater than $\approx 50$), the task remains very difficult, and until the recent advent of deep learning it appeared to be overwhelming, though it can still require the generation of large samples of training data.

It would be very powerful to compare the performance of a given solution to the theoretical upper limit on performance, provided by the true likelihood.  Unfortunately, without knowledge of the true likelihood, it is difficult to assess how well the ML algorithm has captured the necessary information.  For this reason, in the studies presented here and in earlier work~\cite{baldi_searching_2014, baldi_enhanced_2015,Baldi:2016fql}, we built structured datasets with at least two levels of dimensionality: an initial sample with lower-level data at high dimensionality and a reduced sample with expert features at lower dimensionality. Importantly, the expert features are a strict function of the lower-level features, so that they contain a subset of the information. The expertise lies solely in the design of the dimensionality-reducing function, without providing any new information. 

This structure allows us to draw revealing conclusions about the information content of the intermediate and expert-level information and the power of classifiers to extract it. Since the higher-level data contains a subset of the information and benefits from expert knowledge, it can provide the basis for a performance benchmark for the tools using lower-level data in place of the unknown true likelihood. Therefore, if the performance of tools using lower-level data fails to match that of tools using the higher-level data (or a combination of both kinds of data), then we may conclude that the tools using the lower-level data have  failed to extract the complete information.  On the other hand, if the performance of tools using lower-level data exceeds that of tools using the higher-level data, then we may conclude that the higher-level data does not contain all of the information relevant to the classification task, or that it has transformed the problem into a more difficult learning task for the algorithms considered. Regardless of the reason, in this case the transformation to the higher-level lower-dimensional data has failed in its goal.

\section{Data}

Training samples were produced with realistic simulation tools widely used in particle physics. Samples were generated for three classes of jet:

\begin{itemize}
\item light-flavor: jets from $u,d,s$ quarks or gluons;
\item charm: jets from $c$ quarks;
\item heavy-flavor: jets from $b$ quarks.
\end{itemize}

Collisions and immediate decays were generated  with {\sc{madgraph5}}~\cite{madgraph} v2.2.3, showering and hadronization simulated with {\sc pythia}~\cite{pythia} v6.428, and response of the detectors simulated with {\sc delphes}~\cite{delphes} v3.2.0.  Studies with additional $pp$ interactions ({\it pileup}) are reserved for future work; here we assume that pileup effects will not alter the relative performance of the different methods, and is not likely to have a large impact at luminosities recorded to date, given effective techniques to isolate pileup tracks and vertices from the vertices of interest to this study.

The detector simulation was augmented with a simple tracking model that smears truth particles to yield tracks similar to those expected at ATLAS~\cite{Aad:2008zzm}.  Tracks follow helical paths in a perfectly homogeneous $2\,\mathrm{T}$ magnetic field. No attempt was made to account for material interactions or remove strange hadrons. As a result the tracking model lacks the sophistication of models developed by LHC collaborations while retaining enough realism to run vertex reconstruction and compare the relative performance of various machine learning approaches.

Jets are reconstructed from calorimeter energy deposits with the anti-$k_{\mathrm{T}}$ clustering algorithm~\cite{Cacciari:2008gp} as implemented in FastJet~\cite{Cacciari:2011ma}, with a distance parameter of $R = 0.4$. Tracks are assigned to jets by requiring that they be within a cone of $\Delta R \equiv (\Delta \eta^2 + \Delta \phi^2)^{1/2} < 0.4$ of the jet axis.
Jets are labeled by matching to partons within a cone of $\Delta R < 0.5$. If a $b$ or $c$ quark is found within this cone the jet is labeled heavy or charm flavor respectively, with $b$ taking precedence if both are found. Otherwise the jet is labeled light-flavor.

To reconstruct secondary vertices, we use the adaptive vertex reconstruction algorithm implemented in RAVE v6.24~\cite{Waltenberger:2008zza,Waltenberger:2011zz}. The algorithm begins by fitting a primary vertex to the event and removing all compatible tracks. For each jet, secondary vertices are then reconstructed iteratively: a vertex is fit to a point that minimizes $\chi^2$ with respect to all tracks in the jet, less compatible tracks are down-weighted, and the vertex fit is repeated until the fit stabilizes.

Since a $b$-hadron decay typically cascades through a $c$-hadron, jets may include multiple secondary vertices. To account for this, tracks with large weights in the secondary vertex fit are removed and the fit is repeated with the remaining tracks. The process repeats until all tracks are assigned to a secondary vertex.

As described earlier,  we organize the information in three levels of decreasing dimensionality and increasing pre-processing using expert knowledge where each level is a strict function of the lower-level information.   The classification is done per-jet rather than per-event, and at every level the transverse momentum and pseudorapidity of the jet is included.

The lowest-level information considered is the list of reconstructed tracks. Each helical track has five parameters in addition to a $5\times 5$ symmetric covariance matrix with 15 independent entries. The number of tracks varies from 1 to 33 in these samples, with a mean of 4.

The intermediate-level information comes from the output of the vertexing algorithm. The features are the vertex mass, number of tracks associated to the vertex, the fraction of the total energy in jet tracks which is associated to those tracks, vertex displacement, vertex displacement significance, and angular separation in $\Delta \eta$ and $\Delta \phi$ with respect to  the jet axis for each vertex. In cases where both low and intermediate level features are used the track to vertex association weight is also included. The number of vertices varies from 1 to 13 in these samples, with a mean of 1.5.

\begin{figure*}
\begin{center}
\includegraphics[width=0.2\linewidth]{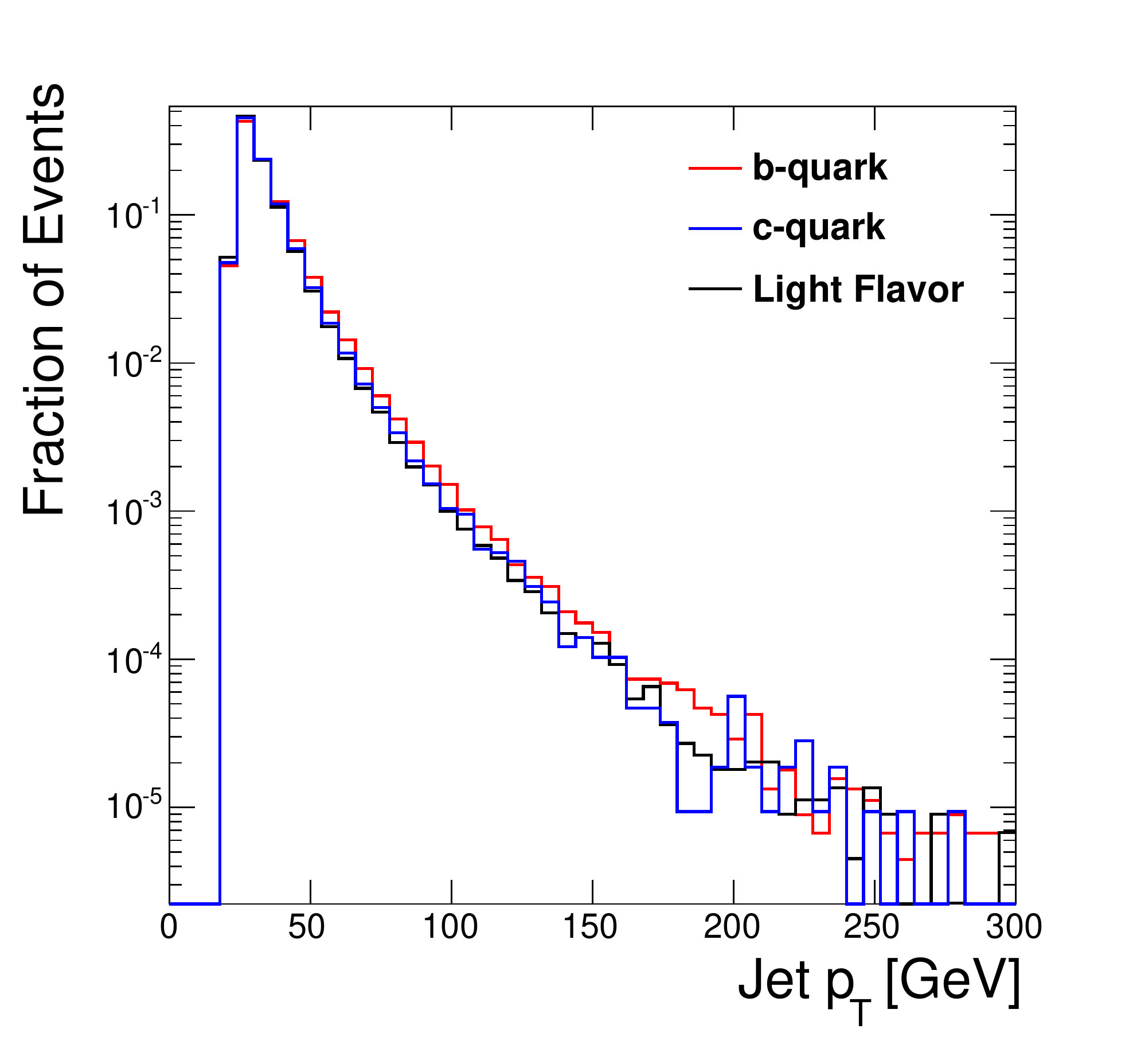}
\includegraphics[width=0.2\linewidth]{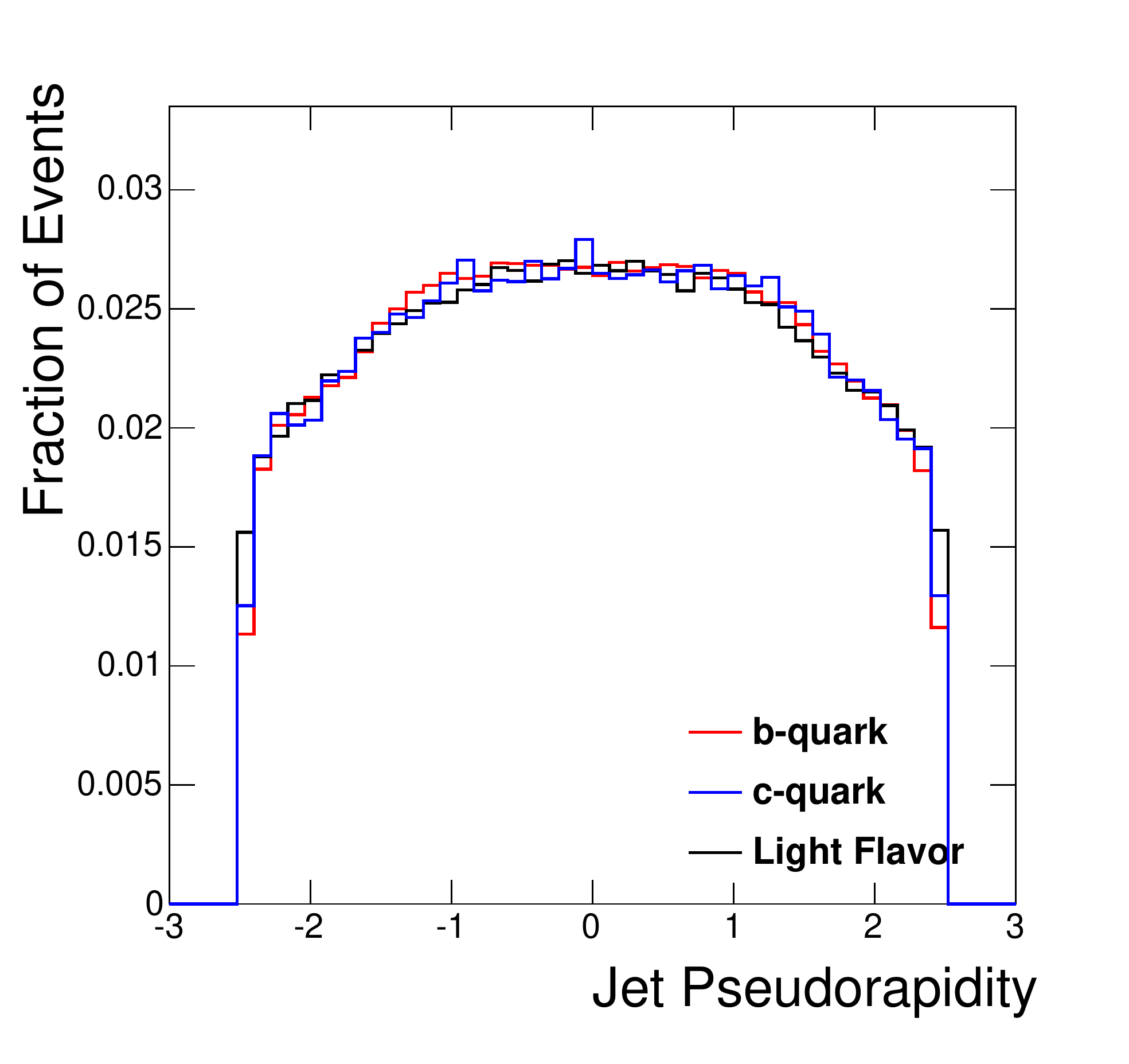}
\includegraphics[width=0.2\linewidth]{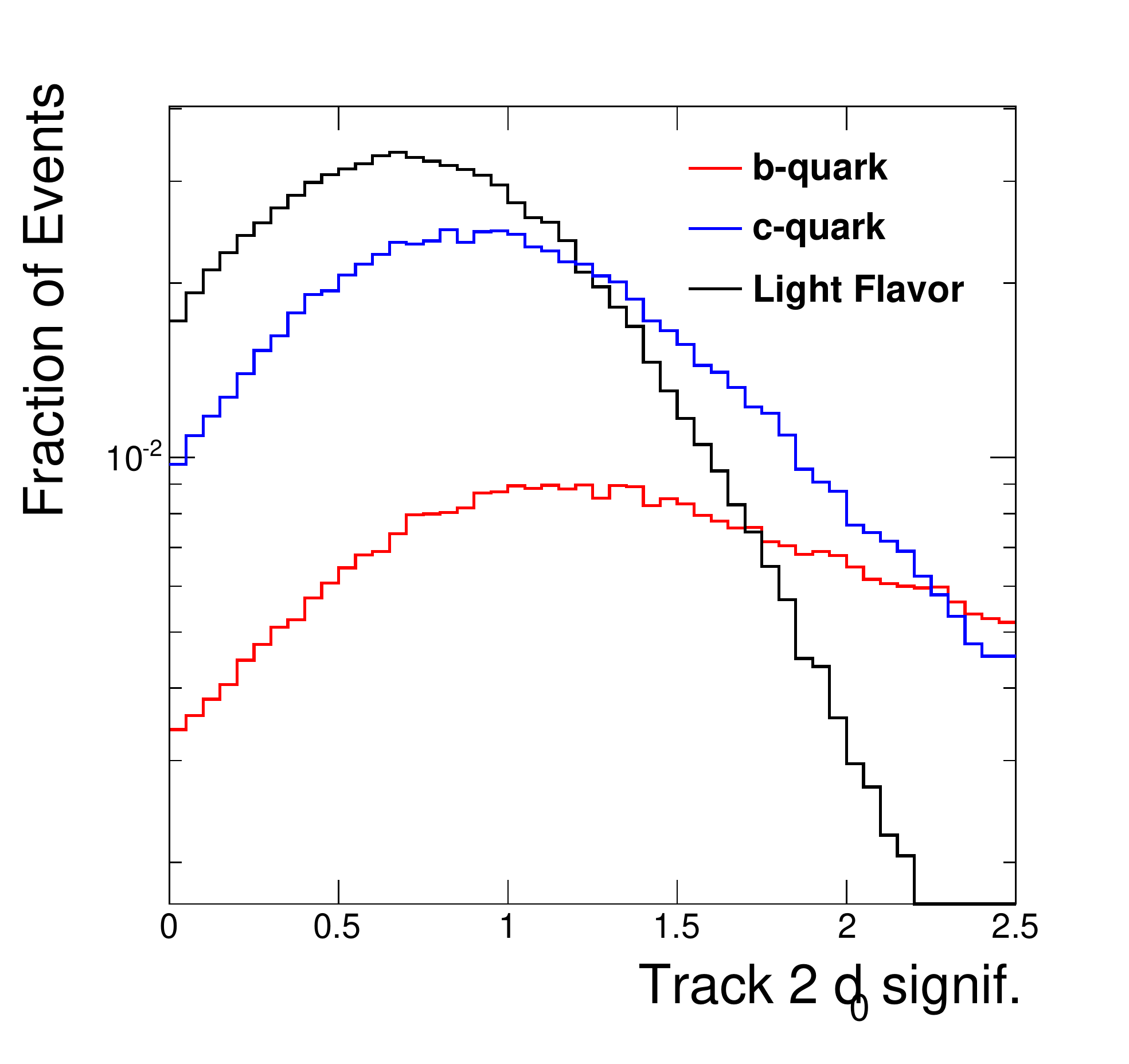}
\includegraphics[width=0.2\linewidth]{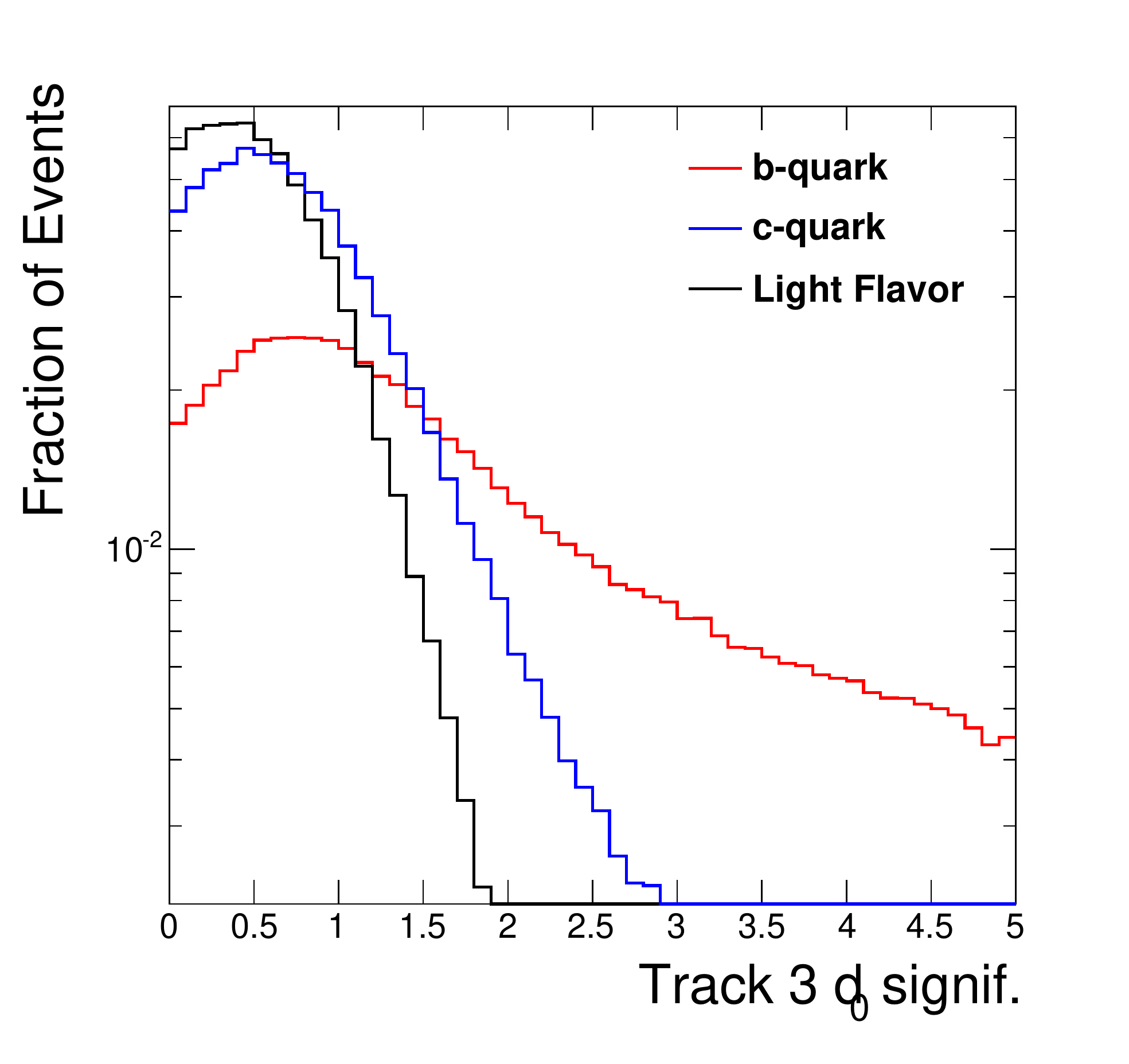}\\
\includegraphics[width=0.2\linewidth]{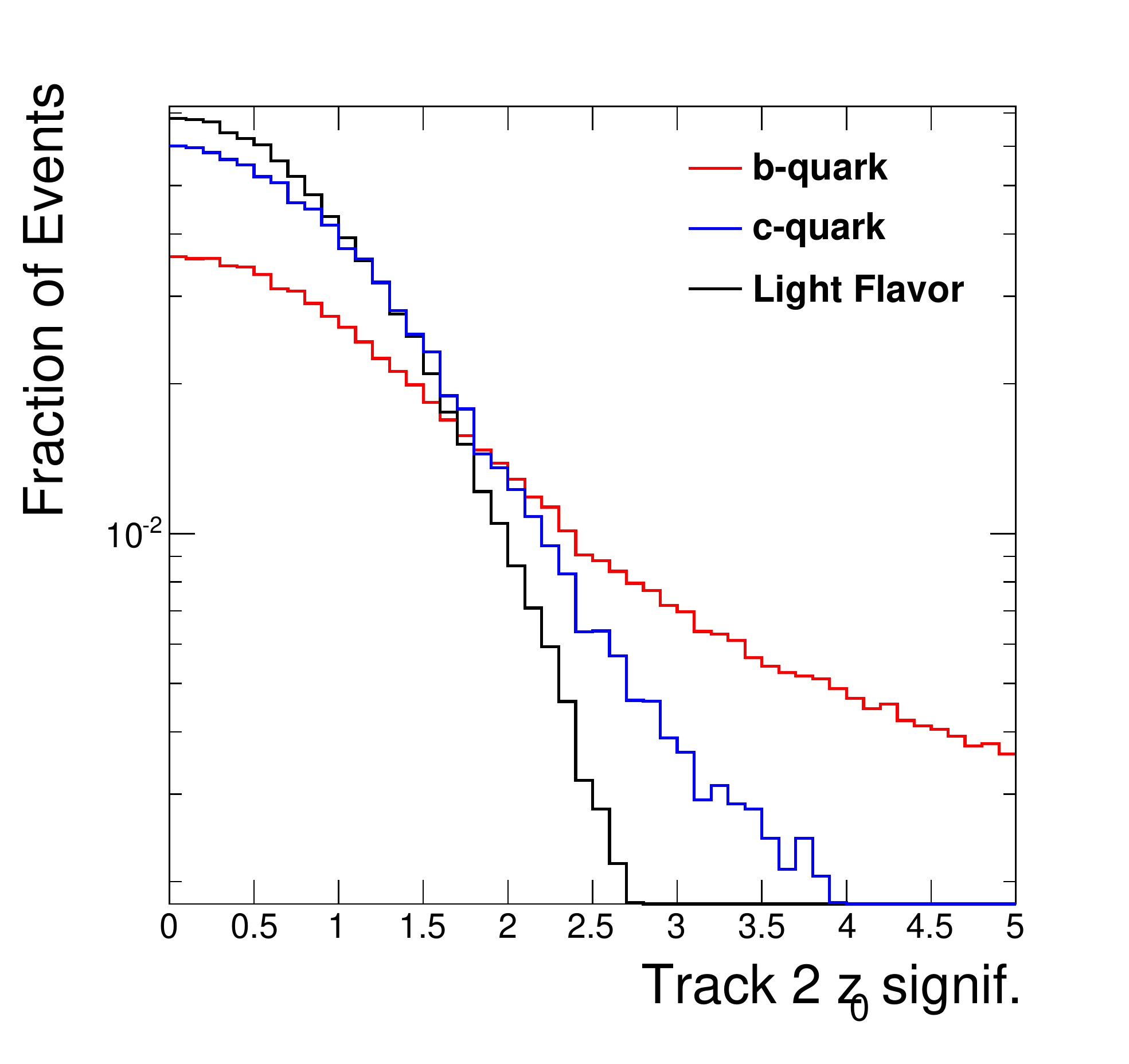}
\includegraphics[width=0.2\linewidth]{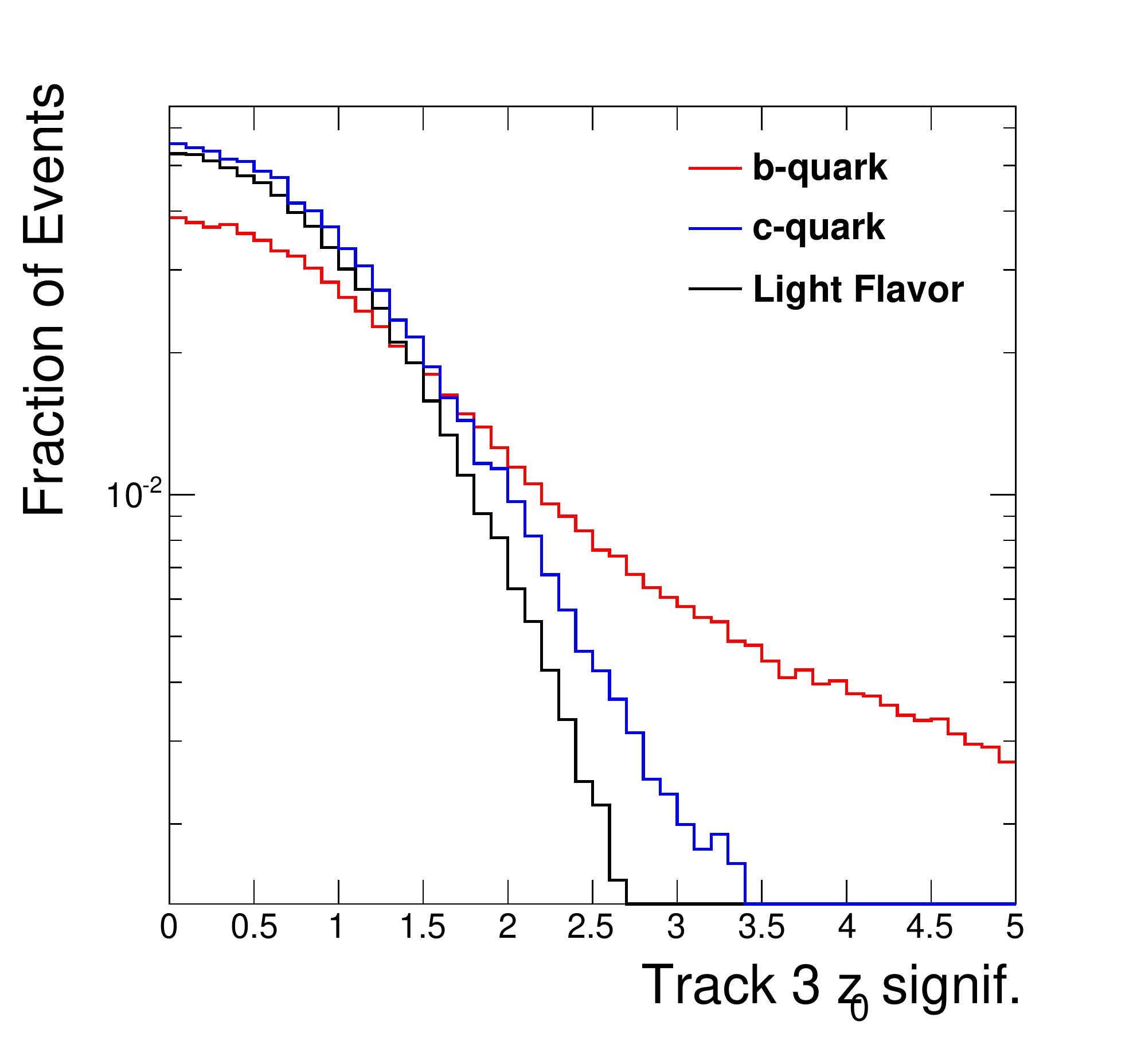}
\includegraphics[width=0.2\linewidth]{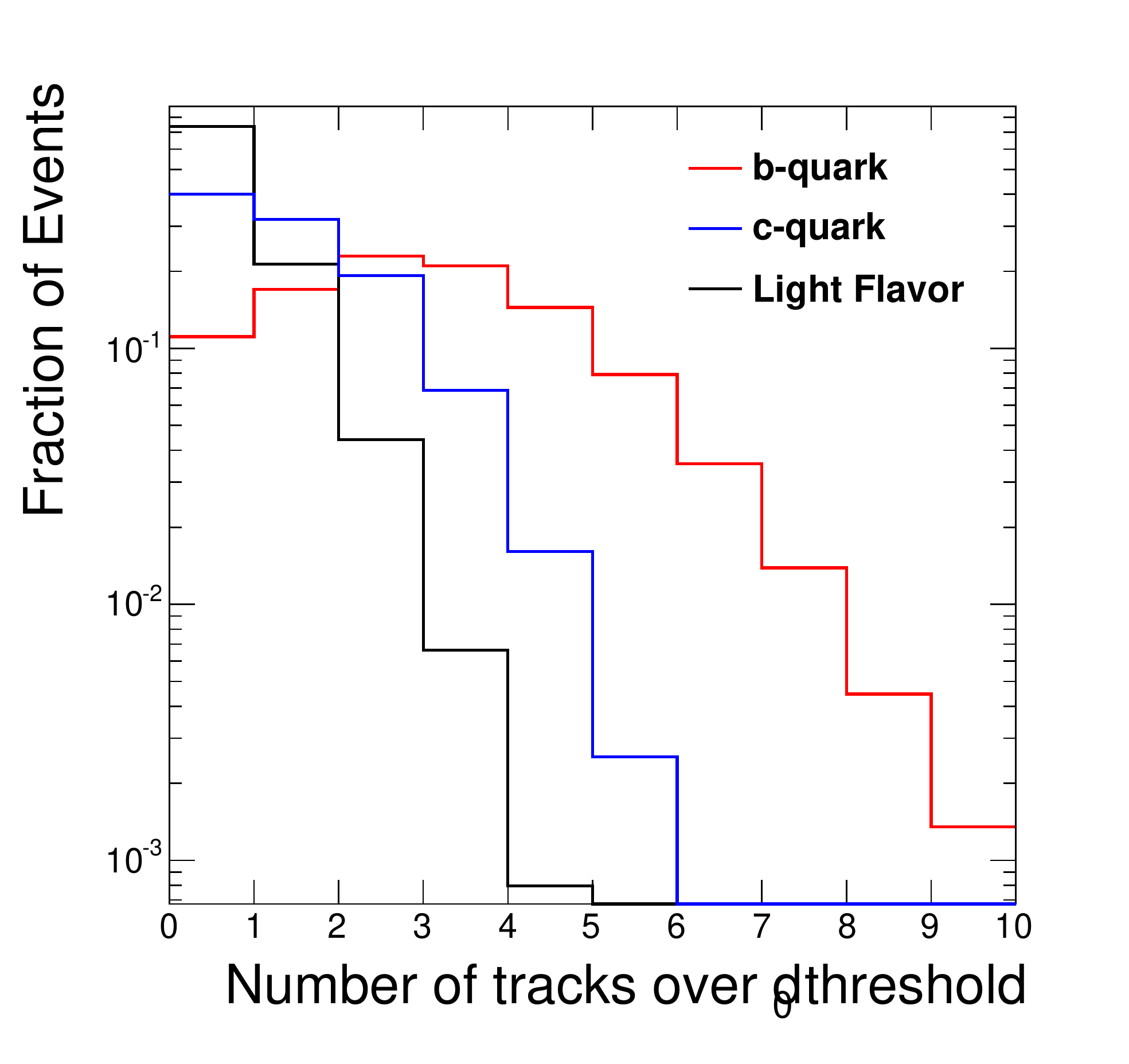}
\includegraphics[width=0.2\linewidth]{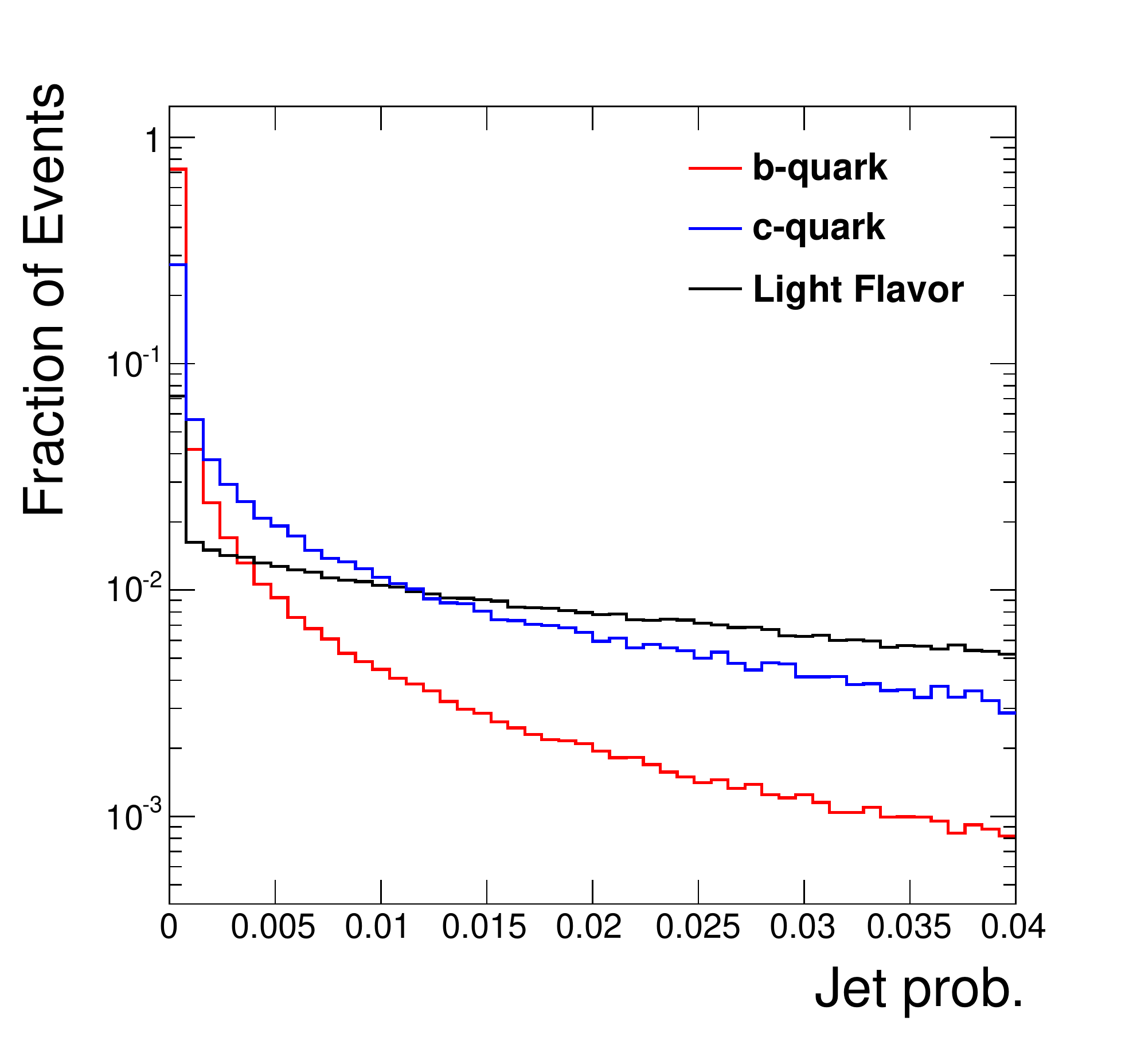}\\
\includegraphics[width=0.2\linewidth]{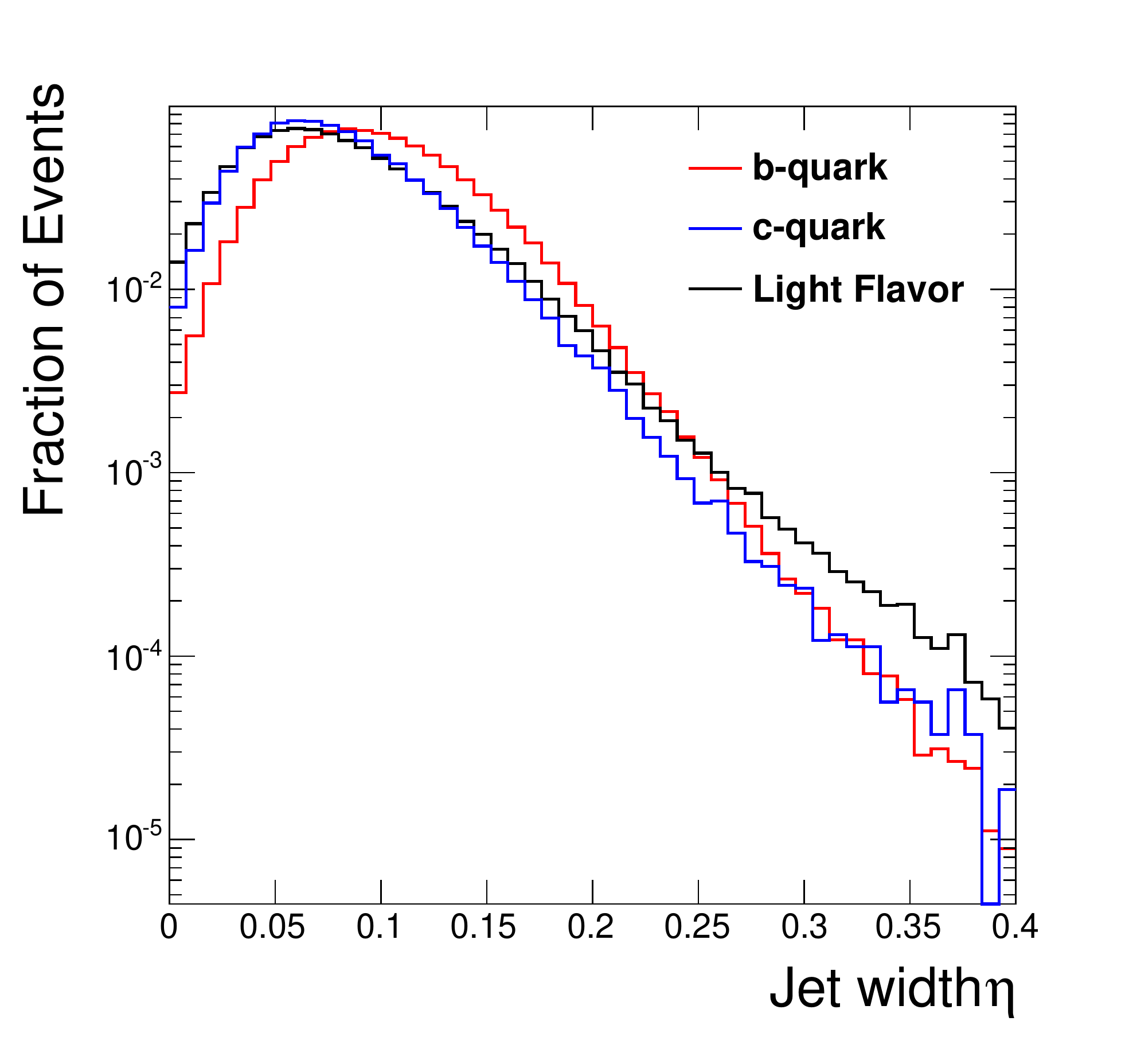}
\includegraphics[width=0.2\linewidth]{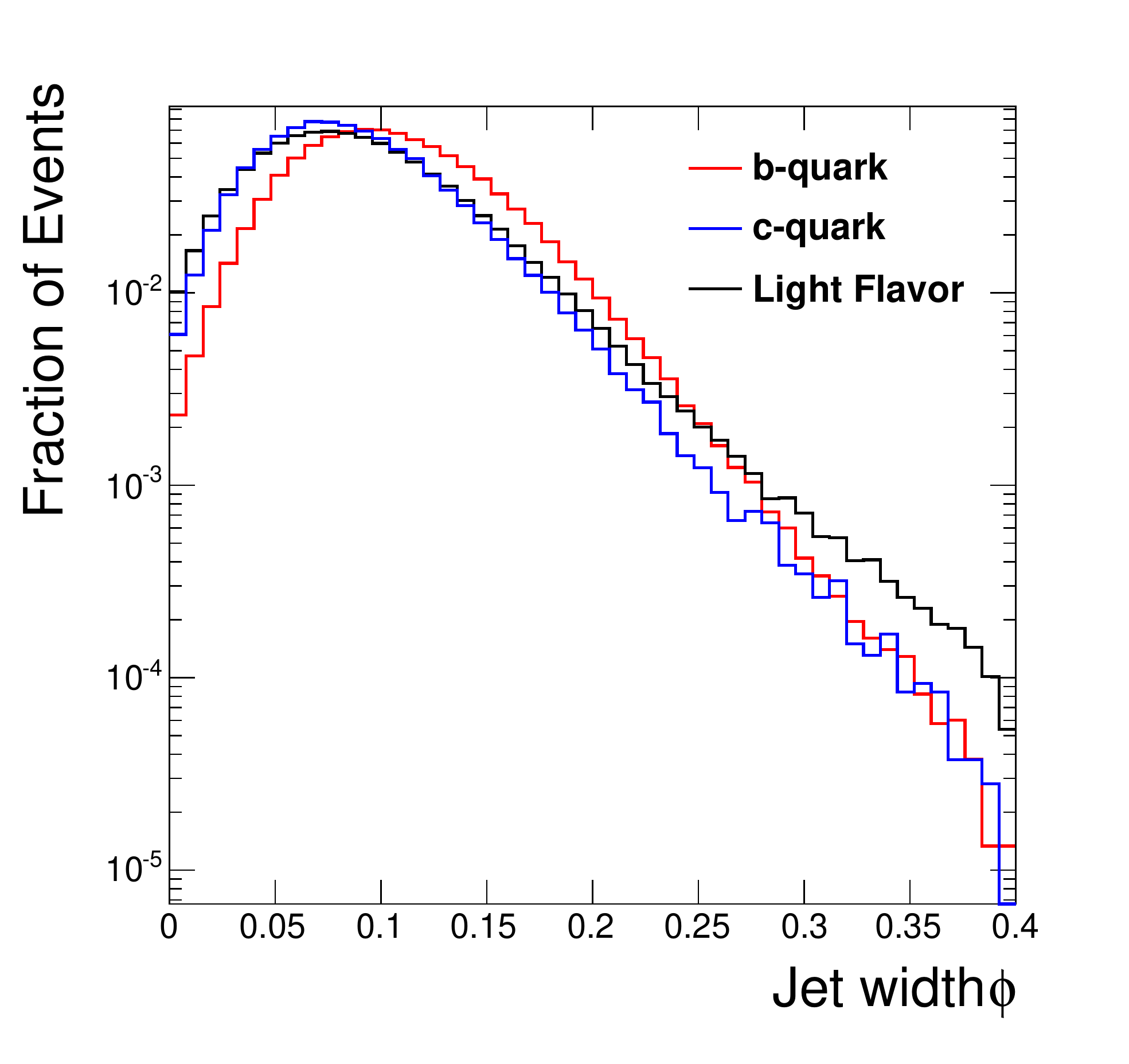}
\includegraphics[width=0.2\linewidth]{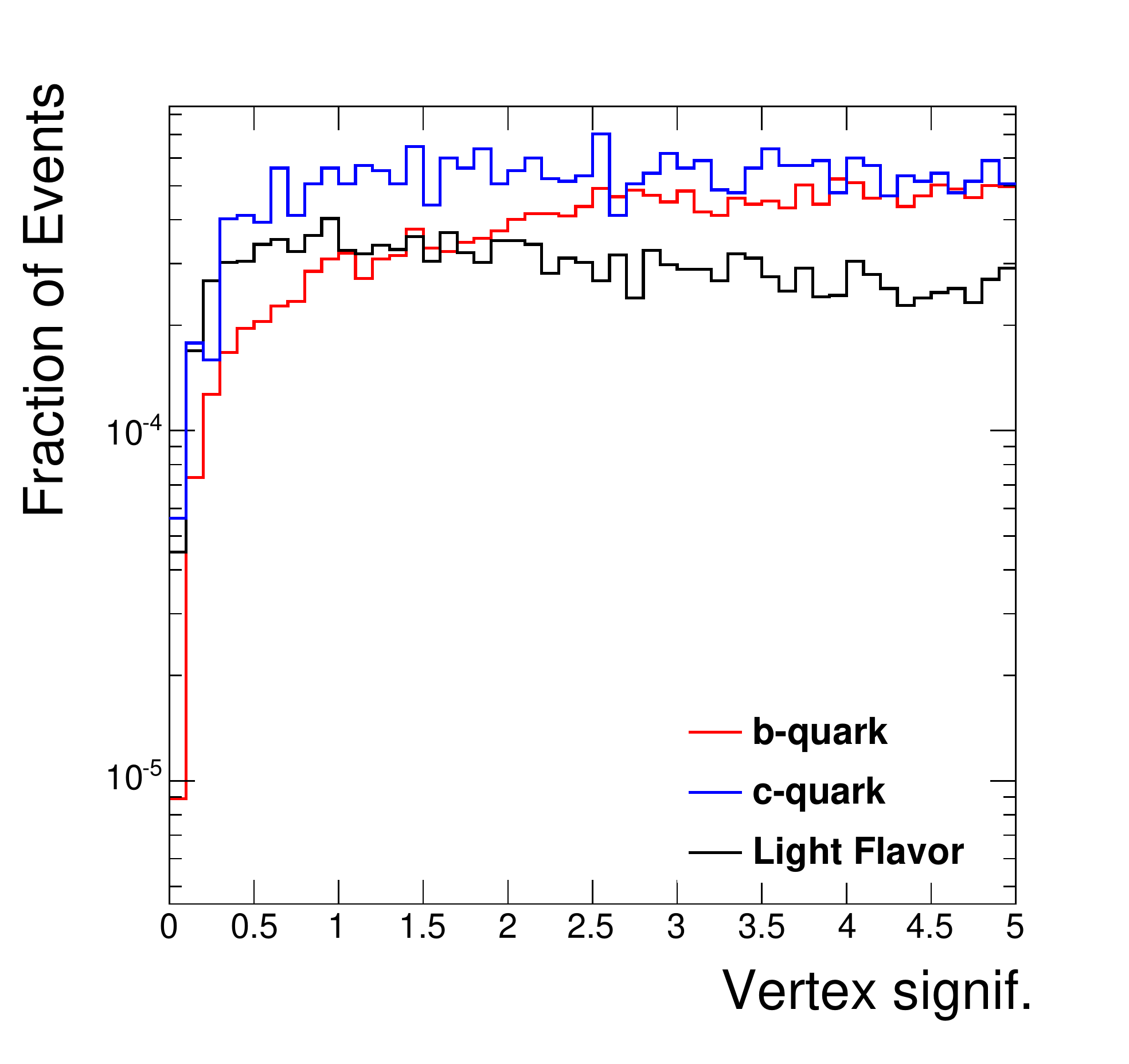}
\includegraphics[width=0.2\linewidth]{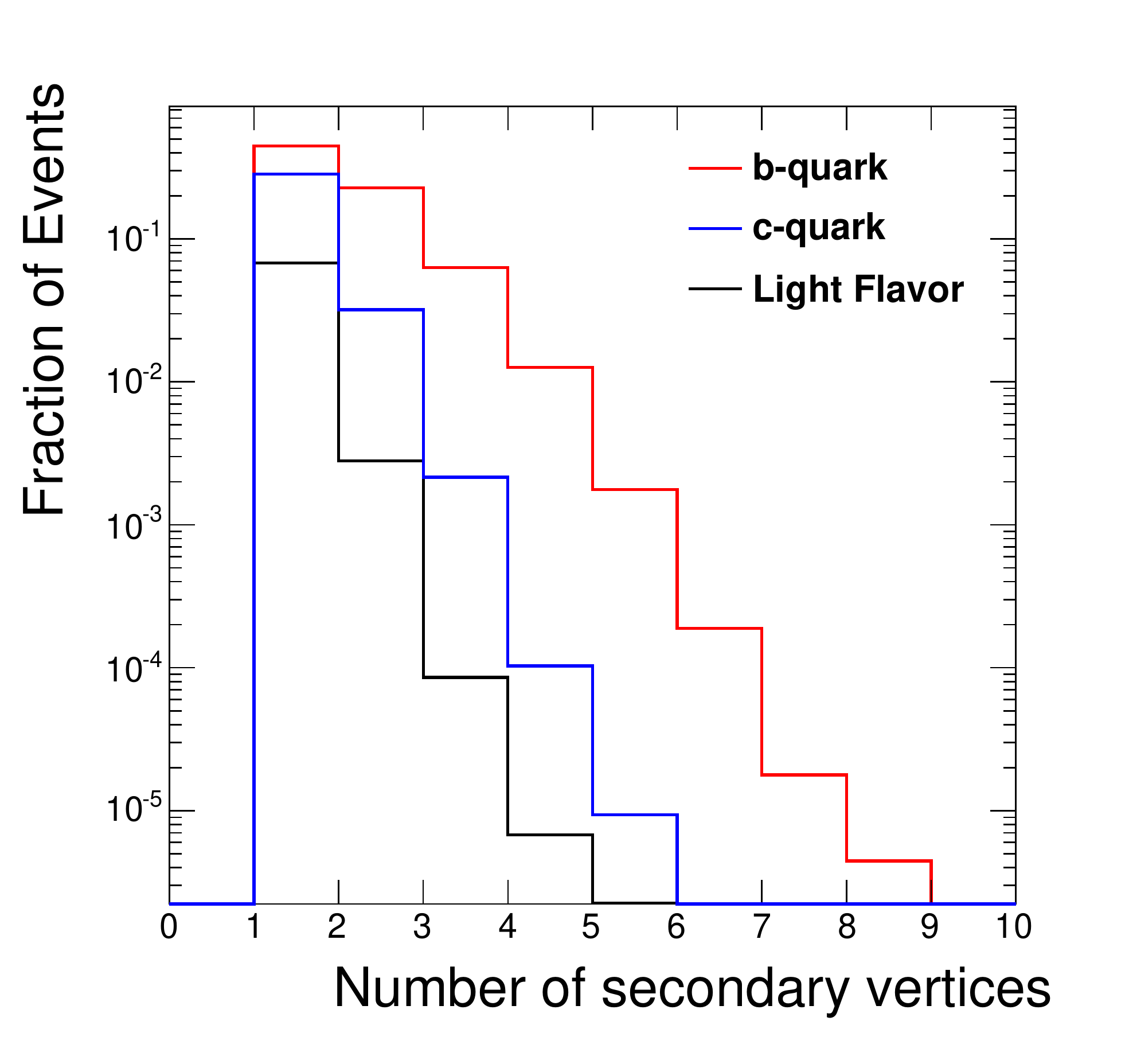}\\
\includegraphics[width=0.2\linewidth]{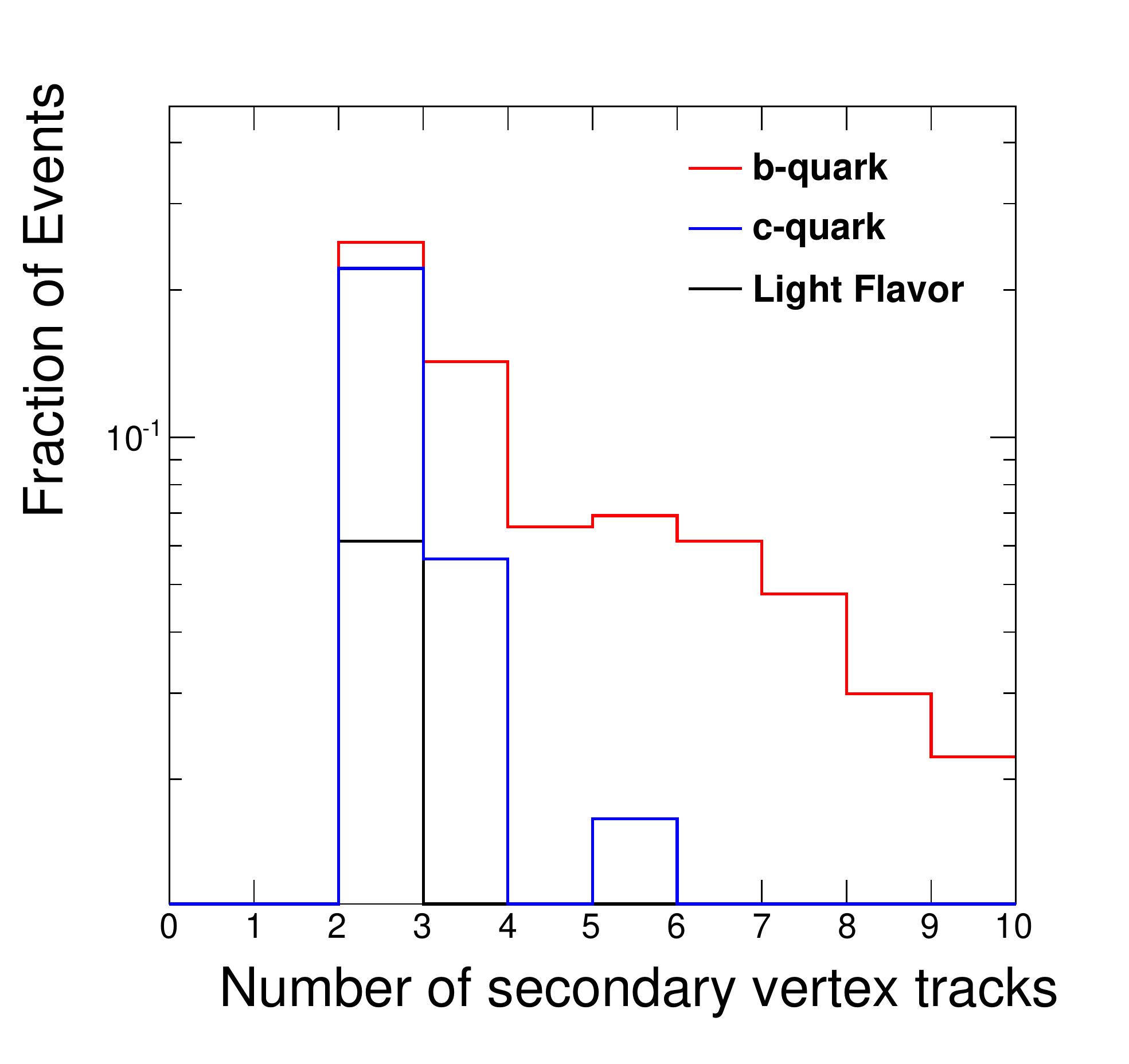}
\includegraphics[width=0.2\linewidth]{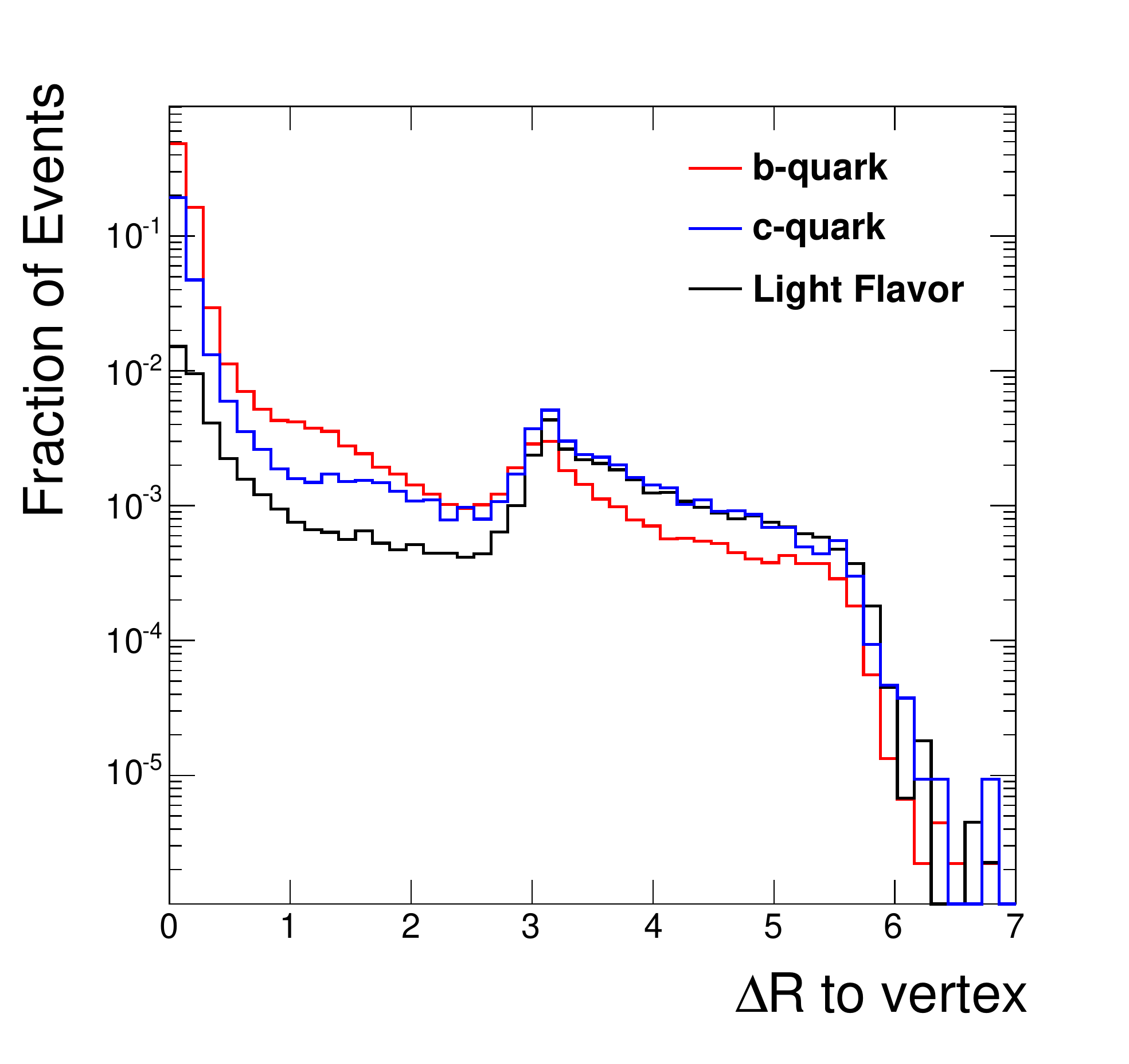}
\includegraphics[width=0.2\linewidth]{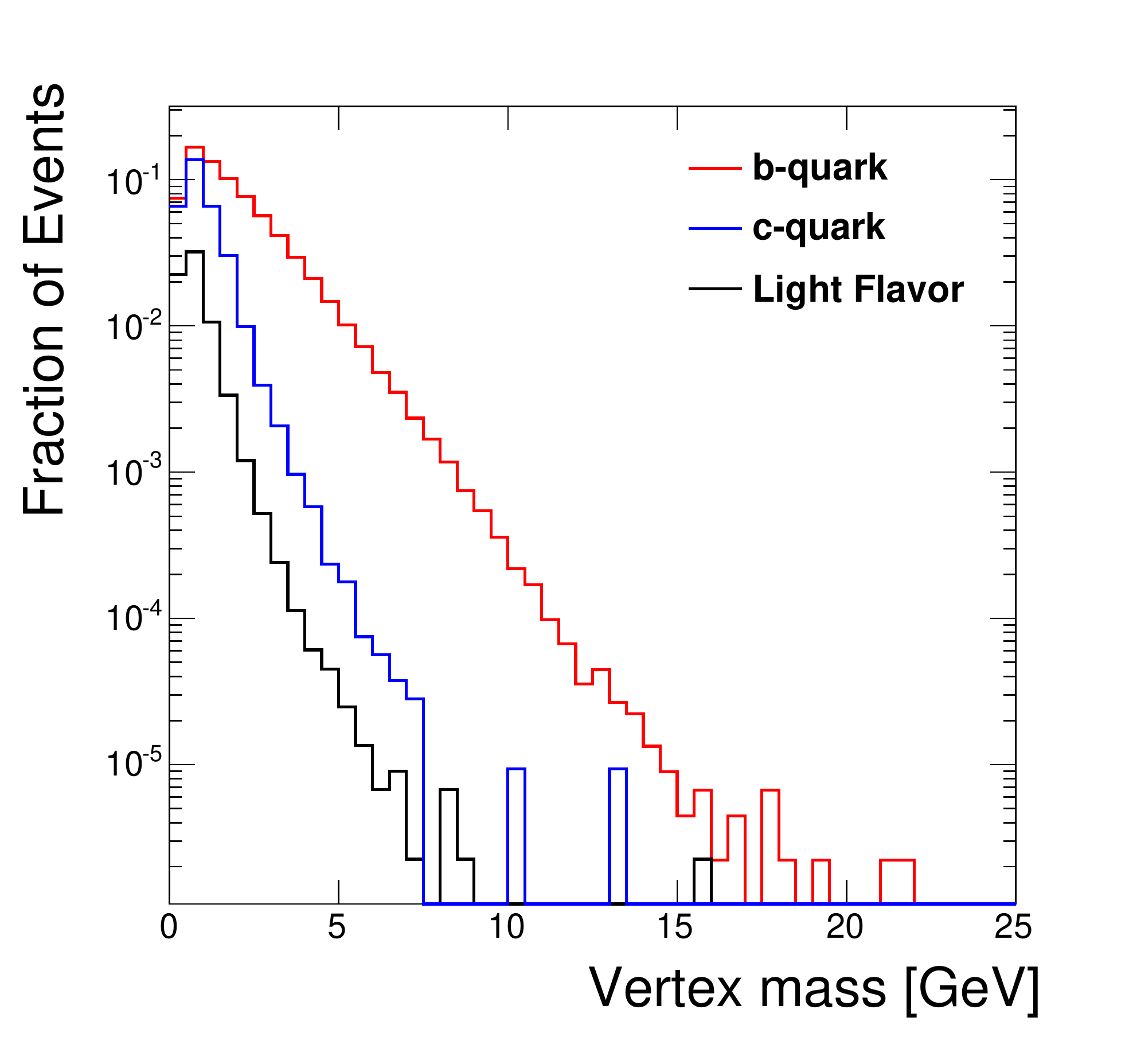}
\includegraphics[width=0.2\linewidth]{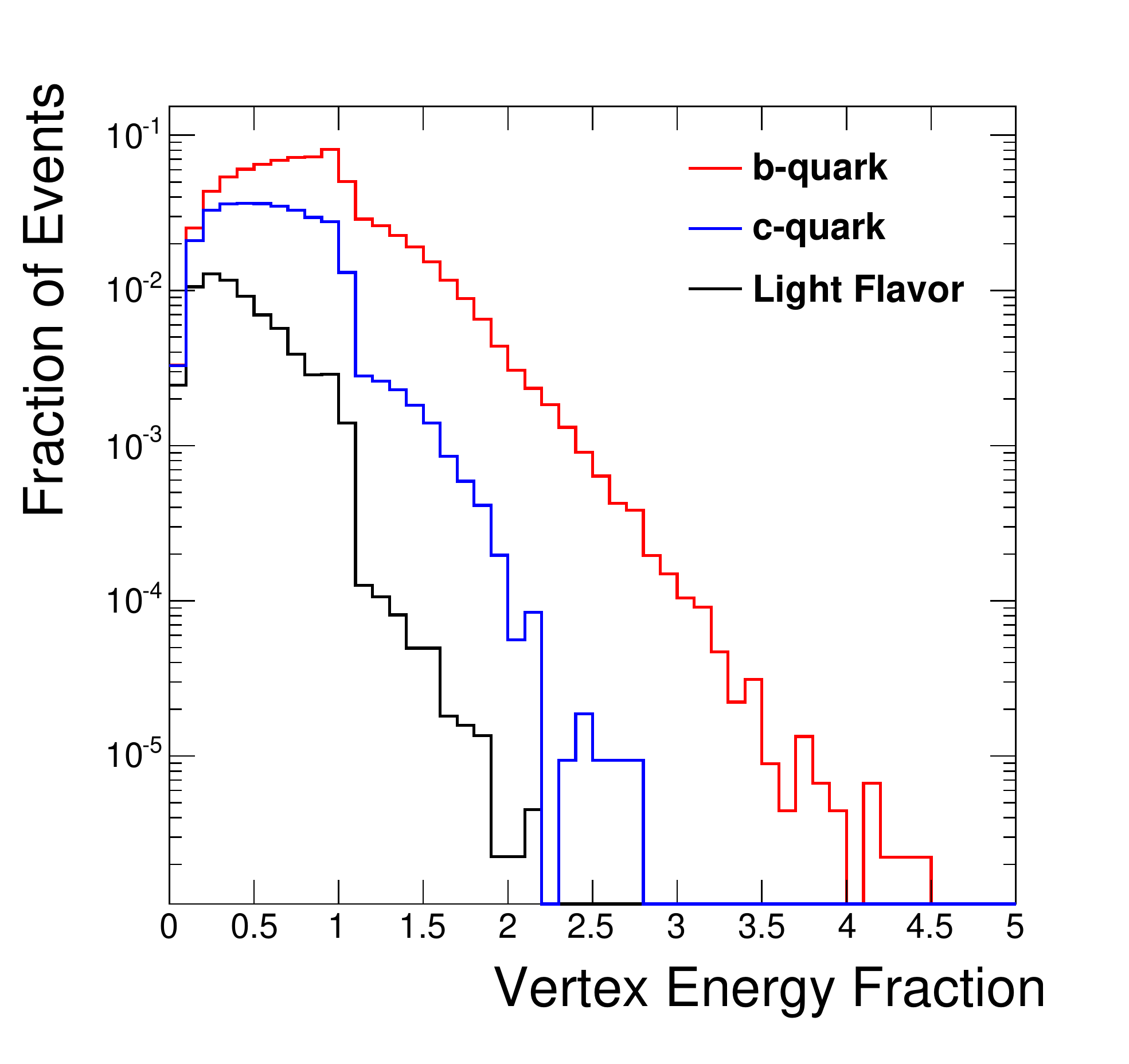}
\end{center}
\caption{Distributions in simulated samples of  high-level jet flavor variables widely used to discriminate between jets from light-flavor and heavy-flavor quarks.}
\label{fig:var}
\end{figure*}

The highest-level information is designed to model the typical features used in current experimental applications; see Fig.~\ref{fig:var} for distributions of these features for each jet class. There are fourteen such features:

\begin{itemize}
\item The $d_0$ and $z_0$ significance of the 2nd and 3rd tracks attached to a vertex, ordered by $d_0$ significance.
\item The number of tracks with  $d_0$ significance greater than $1.8\sigma$.
\item The {\sc JetProb}~\cite{ATLAS-CONF-2010-041} light jet probability, calculated as the product over all tracks in the jet of the probability for a given track to have come from a light-quark jet.
\item The width of the jet in $\eta$ and $\phi$, calculated for $\eta$ as \[ \left( \frac{\sum_{i} p_{\mathrm{T}i} \Delta \eta^2_i}{\sum_i p_{\mathrm{T}}} \right)^{1/2} \] and analogously for $\phi$.
\item The combined vertex significance, \[\frac{\sum_{i} d_i / \sigma_i^2}{ \sqrt{\sum_i 1/\sigma_i^2}}\] where $d$ is the vertex displacement and $\sigma$ is the uncertainty in vertex position along the displacement axis.
\item The number of secondary vertices.
\item The number of secondary-vertex tracks.
\item The angular distance $\Delta R$ between the jet and vertex.
\item The decay chain mass, calculated as the sum of the invariant masses of all reconstructed vertices, where particles are assigned the pion mass.
\item The fraction of the total track energy in the jet associated to secondary vertices \footnote{ The vertex energy fraction is not a strict fraction; it can be greater than unity if tracks are assigned to multiple vertices.}
\end{itemize}

The dataset consists of 10 million labeled simulated jets.  The corresponding target labels are “light-flavor”, “charm”, and “heavy-flavor”. The data contains 44, 11, 45 percent of each class respectively.  This data is available from the UCI Machine Learning in Physics Web portal at \url{http://mlphysics.ics.uci.edu/}.

\section{Methods}

In the experiments, we typically use 8 million samples for training, one million for validation, and one million for testing.  Since there are three labels but we are interested in the study of signal vs background and classification, the labels are converted to binary by mapping bottom quark to one, and both charm and light quark to zero. We study the light-quark and charm-quark rejection separately.

\begin{figure}%
    \begin{center}
    \includegraphics[width=0.8\linewidth]{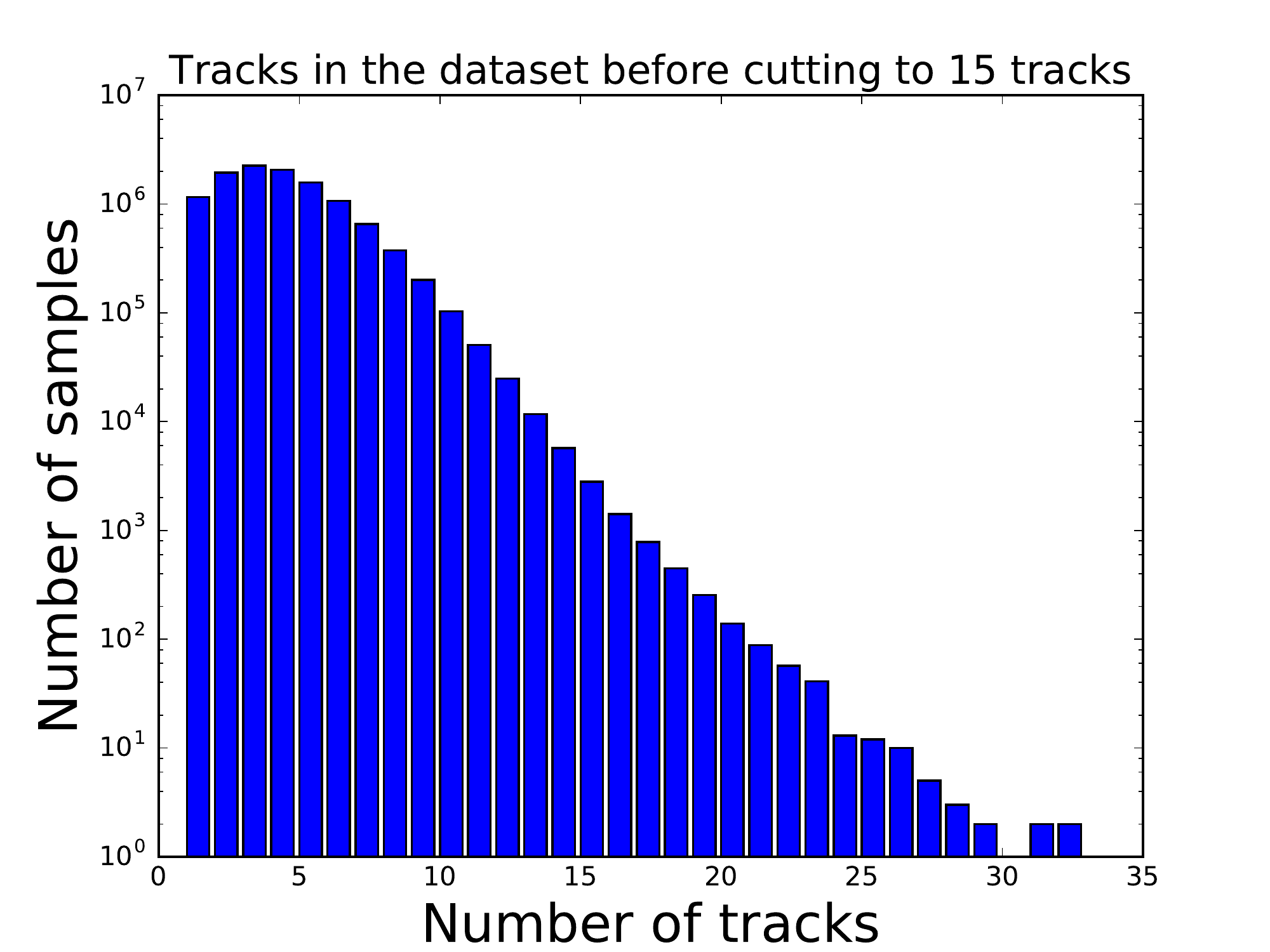}\\
    \includegraphics[width=0.8\linewidth]{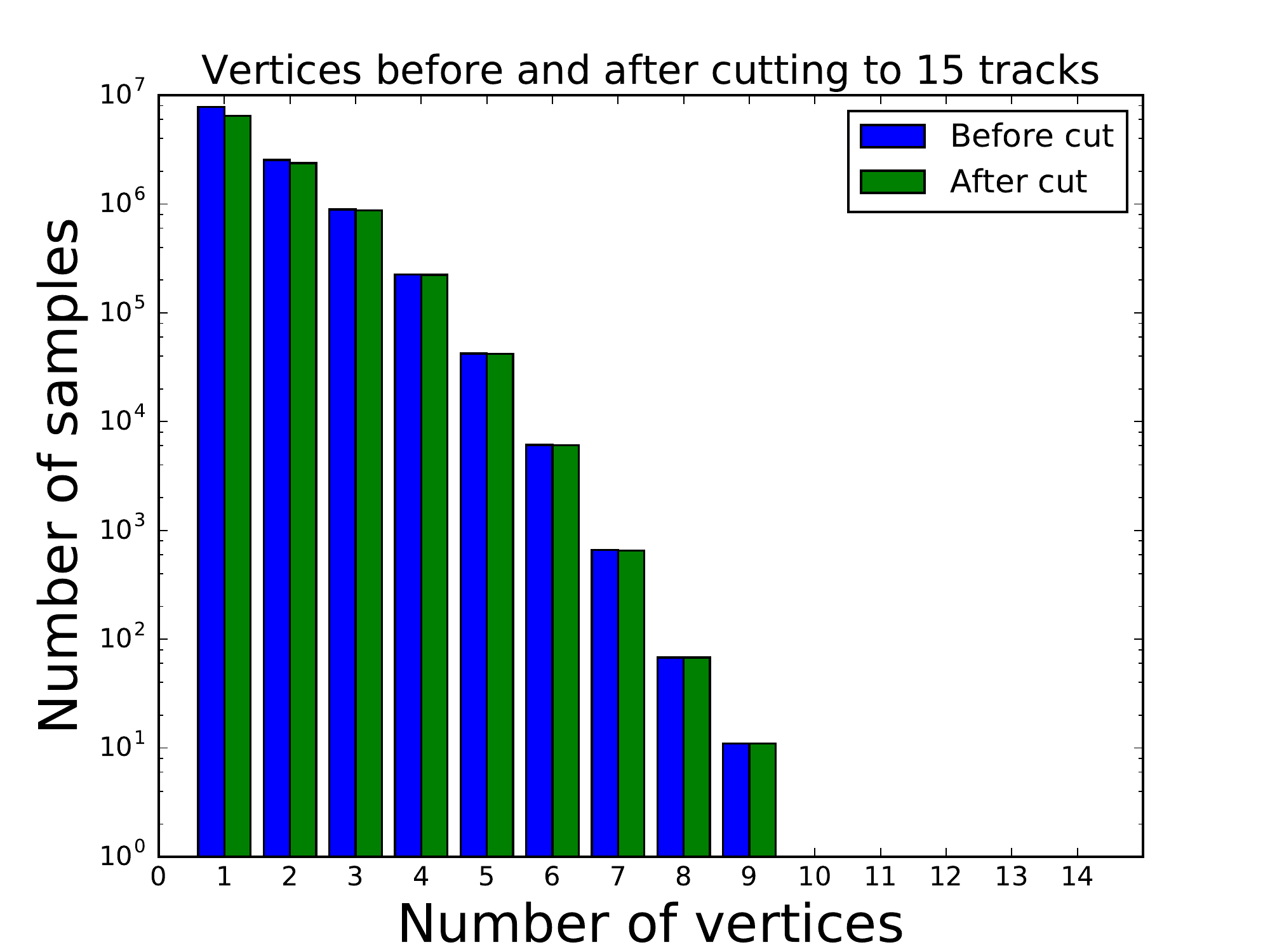}
    \caption{Top: Distribution of the number of tracks associated to a jet in simulated samples. Bottom: Distribution of the number of vertices associated to a jet in simulated samples, before and after removing tracks which exceed the maximum allowed value of 15.}
    \label{fig:track-vertex-dist}
    \end{center}
\end{figure}

\subsection{Machine Learning Approaches}

To each simulated collision is attached a set of tracks and a set of vertices. This poses  challenges for a machine learning approach in that the size of these sets is variable as seen in Fig.~\ref{fig:track-vertex-dist} and the sets are unordered, although as usual an arbitrary order is often used to list their elements. To address and explore these challenges we use three different deep learning approaches: feedforward neural networks, recurrent neural networks with LSTM (Long Short Term Memory) units, and outer recursive neural networks.

\subsubsection{Feedforward Neural Networks}

The track feature set and the vertex feature set have variable size for a given collision. However, the structure of  feedforward networks requires a fixed-size input to make predictions.
Thus the use of feedforward neural networks requires first an arbitrary ordering and then a capping of the size of the input set, with zero padding for sets that are smaller than the capped size. To resolve the arbitrary ordering the tracks were sorted by decreasing absolute $d_0$ significance. This ordering also ensures that tracks from a secondary vertex, which typically have large $d_0$, are unlikely to be removed by the capping.
Random ordering before adding the padding was also tested but the performance was lower than using the absolute $d_0$ significance ordering.

To create a fixed size input, the number of tracks was limited to 15, from a maximum of 33. Using 15 as the cutoff value ensures that 99.97\% of the samples preserve all their original tracks; see Fig.~\ref{fig:track-vertex-dist}.
Tracks are associated to vertices by concatenating the track parameters with those from the associated vertex.
Before training, the samples are preprocessed by shifting and scaling such that each feature has a mean of zero and a standard deviation of one. Jets with fewer than 15 tracks are zero-padded after preprocessing. After the cut on the number of tracks, the maximum number of vertices is 12 with an average of 1.5; see Fig.~\ref{fig:track-vertex-dist}.

The feedforward neural networks were trained on 8 million training samples with one million more for validation using stochastic gradient descent with mini-batches of 100 samples. They were trained for 100 epochs and the best model was chosen based on the validation error. Momentum for the weights updated was used and linearly increased from zero to a final value over a specified number of epochs. Learning rate decayed linearly from 0.01 to a final value starting and finishing at a specified number of epochs. Dropout (in which nodes are removed during training) with values of $p$ from 0.0 to 0.5 were used at several combinations of layers to add regularization \cite{hinton_improving_2012, baldi_dropout_2014}. These networks had 9 fully connected hidden layers with rectified linear units \cite{glorot_deep_2011, jarrett_what_2009}.

\begin{figure}%
    \begin{center}
    \includegraphics[width=1\linewidth]{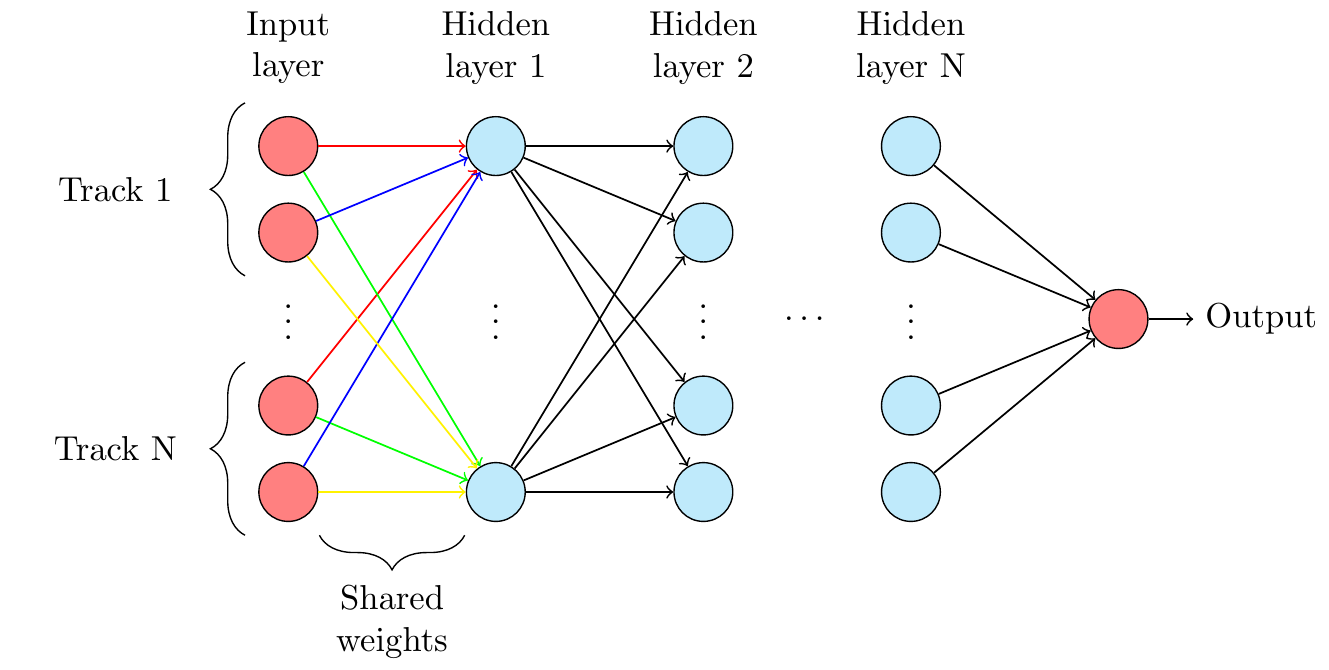}
    \caption{Feedforward neural network architecture. In the first layer, connections of the same color represent the same value of the shared weight. The others layers are fully connected without shared weights.}
    \label{fig:ff-NN-diagram}
    \end{center}
\end{figure}

Shared weights for each track object were used at the first layer  to preserve information about the structure of the data; see Fig~\ref{fig:ff-NN-diagram}. When adding the vertex and high level variables to the tracks, these were also included within the set of variables with shared weights.  The weights for all but the last layer were initialized from a uniform distribution between $[-\sqrt{6/C}, \sqrt{6/C}]$ where $C$ is the total number of incoming and outgoing connections~\cite{glorot_weight_init}. The weights for the last layer were initialized from a uniform distribution between -0.05 and 0.05. A manual optimization was performed over all the hyperparameters to find the best model.

\subsubsection{LSTM Networks}

\begin{figure}
    \begin{center}
    \includegraphics[width=1\linewidth]{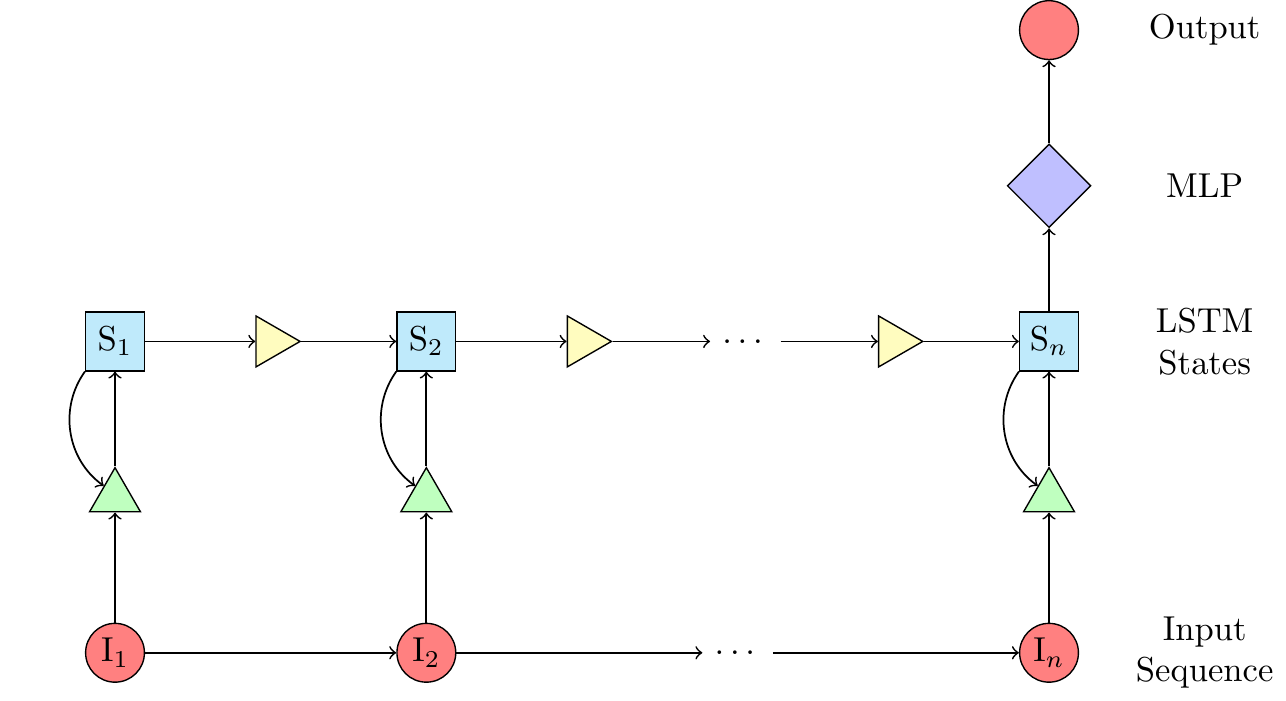}
    \caption{Architecture of the Long Short Term Memory networks as described in the text.}
    \label{fig:LSTM-diagram}
    \end{center}
\end{figure}

A natural approach to handling variable-sized input is to use recursive neural networks. Broadly speaking, there are two classes of approaches for designing such architectures, the inner approach and the outer approach~\cite{baldi2016innerouter}. In the inner approach, neural networks are used inside the data graphs to crawl the corresponding edges and compute the final output. This process requires the data graphs to be directed and acyclic. Since here the data consists of a set of vertices and tracks, we first convert the data into a sequence by ordering the vertices and tracks as described previously and then use recursive neural networks for sequences, in combination with Long Short Term Memory units \cite{gers2000learning, greff2015lstm} to better capture long range dependencies. In the underlying acyclic graph, the variables associated with each node are a function of the variables associated with the parent nodes. Each such function can be parameterized by a neural network. Because the directed acyclic graph has a regular structure, the same network can be applied at different locations of the graph, ultimately producing the LSTM grid network in Figure \ref{fig:LSTM-diagram}. 

We follow the standard implementation of LSTMs with three gates (input, forget, output) and initialize the connections to random orthonormal matrices. The input data consists of a sequence of concatenated track, vertex, and expert features (or different sub-combinations thereof) which are sorted by their absolute $d_0$ significance, as was the case with the fully connected models. The main difference is that we do not need zero-padding as the LSTM networks can handle sequences of arbitrary length, though we retain the same maximum of 15 tracks for comparability. The final model consists of one LSTM layer comprising between 50 and 500 neurons, and a feedforward neural network with one to four hidden layers that receives its input from the LSTM network and produces the final predictions (where each layer has between 50 and 500 units). We add dropout layers in between the LSTM and each hidden fully connected layer. For hyperparameter-optimization we performed a random search over these parameters as well as the individual dropout rates that are part of the model. We trained the LSTM networks for 100 epochs using SGD with a momentum of 0.9 and decay the step-size parameter from initially $2 \cdot 10^{-3}$ down to $10^{-4}$ over the course of training.

\subsubsection{Outer Recursive Networks}

\begin{figure}
    \begin{center}
    \includegraphics[width=1\linewidth]{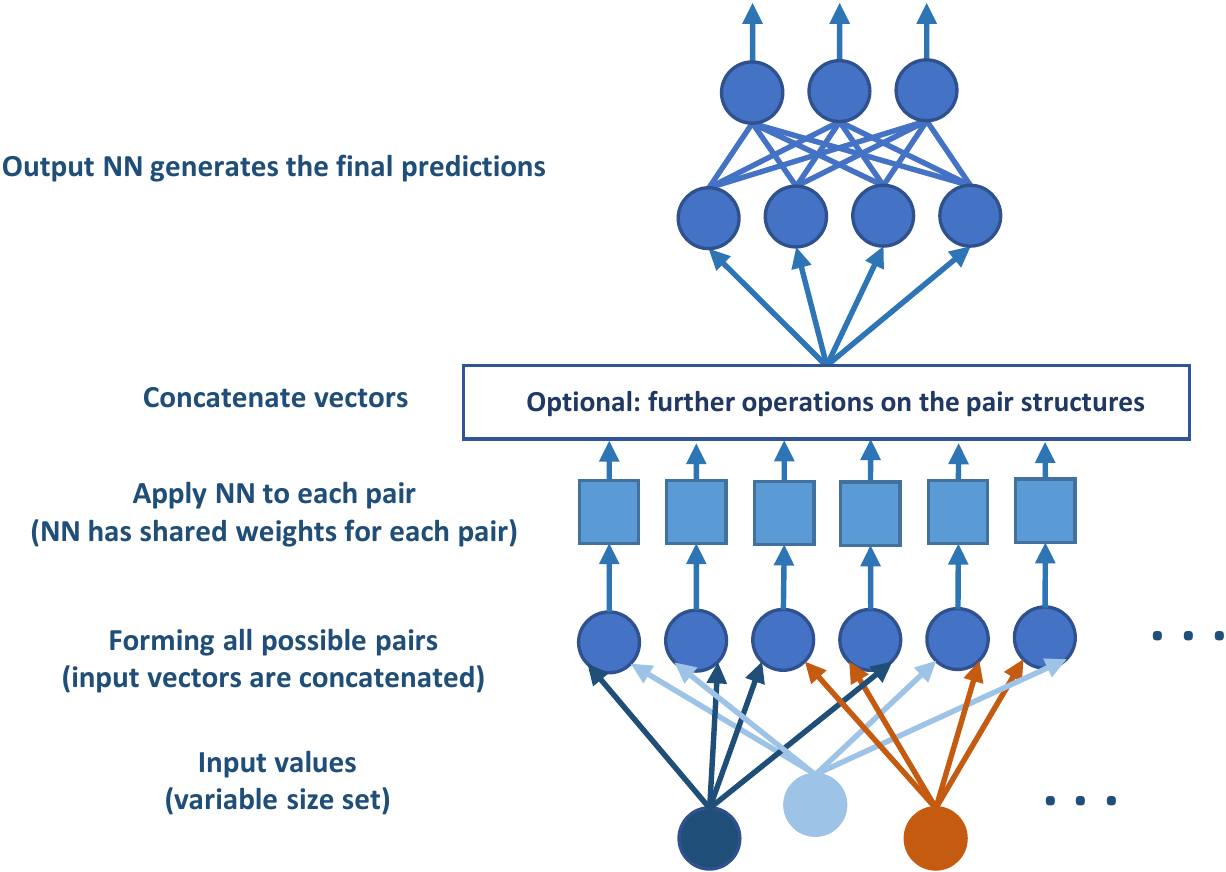}
    \caption{Architecture of the outer recursive networks as described in the text.}
    \label{fig:outer-recursive-diagram}
    \end{center}
\end{figure}

Alternatively, to handle inputs of variable size, we can use an outer recursive approach, where neural networks are built in a direction perpendicular to the original data graph, with horizontal weight sharing. The outer approach can be used to build more symmetric deep architectures; see Fig.~\ref{fig:outer-recursive-diagram}. For instance, in our case the input consists of up to 15 tracks, from which we can sample all possible pairs of tracks and use a shared neural network that processes these in the first layer of the outer approach. In this case, there are at most ${15 \choose 2} = 105$ unordered pairs, or 210 ordered pairs, which is manageable especially considering that there is a single network shared by all pairs. Using ordered pairs would yield the most symmetric overall network. At the next level of the architecture, one can for instance use a network for each track $t_i$ that combines the outputs of all the networks from the first layer associated with pairs containing $t_i$, and so forth. In the second level of the outer architecture, for simplicity here we use a fully connected feedforward network that computes the final output using the outputs of all the pair networks. More specifically, for each data sample we compute the list of stacked track features for all 210 pairs and process each pair with a shared nonlinear hidden layer (with 5 to 20 neurons). The resulting  outputs for all pairs are then concatenated and fed into a multilayer perceptron as was the case for the LSTM models, with one to four hidden layers containing between 100 and 600 hidden units. We again use dropout layers in between the hidden layers and optimize the dropout rates and network depth and size using random search.

\subsection{Hardware and Software Implementations} 

All computations were performed using machines with 16 Intel Xeon cores, NVIDIA Titan graphics processors, and 64 GB memory. All neural networks were trained using the GPU-accelerated Theano software library \cite{theano_2016} and, for the feed forward neural networks, also the Keras software library \cite{chollet_keras_2015}.

\section{Results}

The best feedforward neural networks have 9 fully connected hidden layers with 400 rectified linear units and a  single sigmoid unit at the end. On the first layer the networks have shared weights. The first five tracks have one set of shared weights per track, tracks 6 to 10 have a second set of shared weights per track and the last five tracks have a third set of shared weights per track. They have a momentum term of 0 which starts to linearly increase at the first epoch and reaches its final value of 0.5 at epoch 100. Initially, the learning rate is set at 0.01 and, starting at epoch 80, it is linearly  decreased to a final value of 0.001 at epoch 100. Dropout was used in the first two layers with a value of p=0.3. The same architecture was used across all the combinations of features except in the case of using only high level features, in which case the first layer is fully connected without any shared weights. 

We found that the main characteristic of the best LSTM models is a relatively small size of the hidden state representation of the LSTM module (about 70 units), while the size of the MLP, which is sitting on top of it, is of secondary importance for overall performance of the model. The best models using the outer recursive approach contain between two and three hidden layers on top of the shared-weight layer (which operates on all paired tracks) and those contain 17 or more neurons.

\begin{figure*}
\begin{center}
\includegraphics[width=0.7\linewidth]{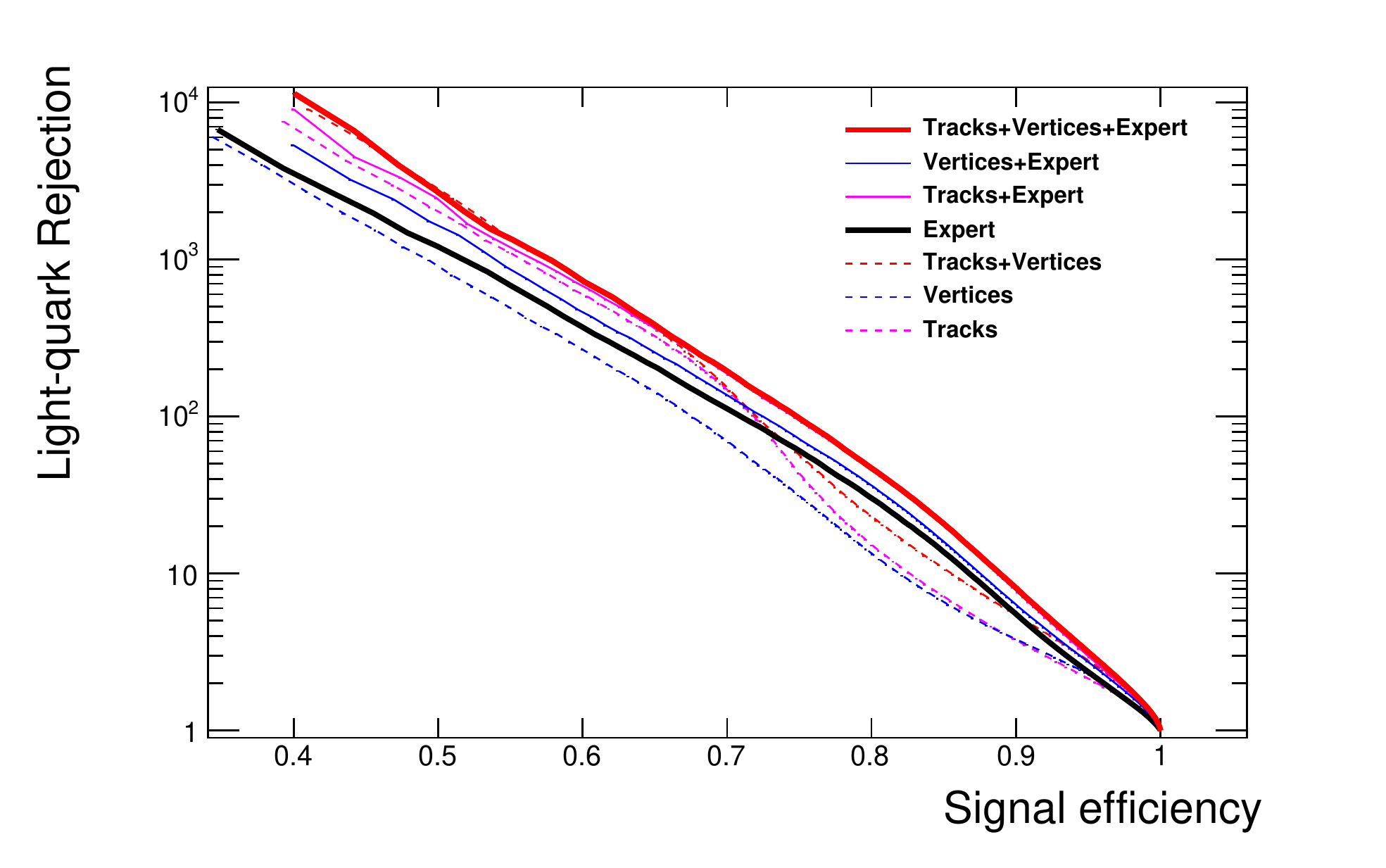}
\includegraphics[width=0.7\linewidth]{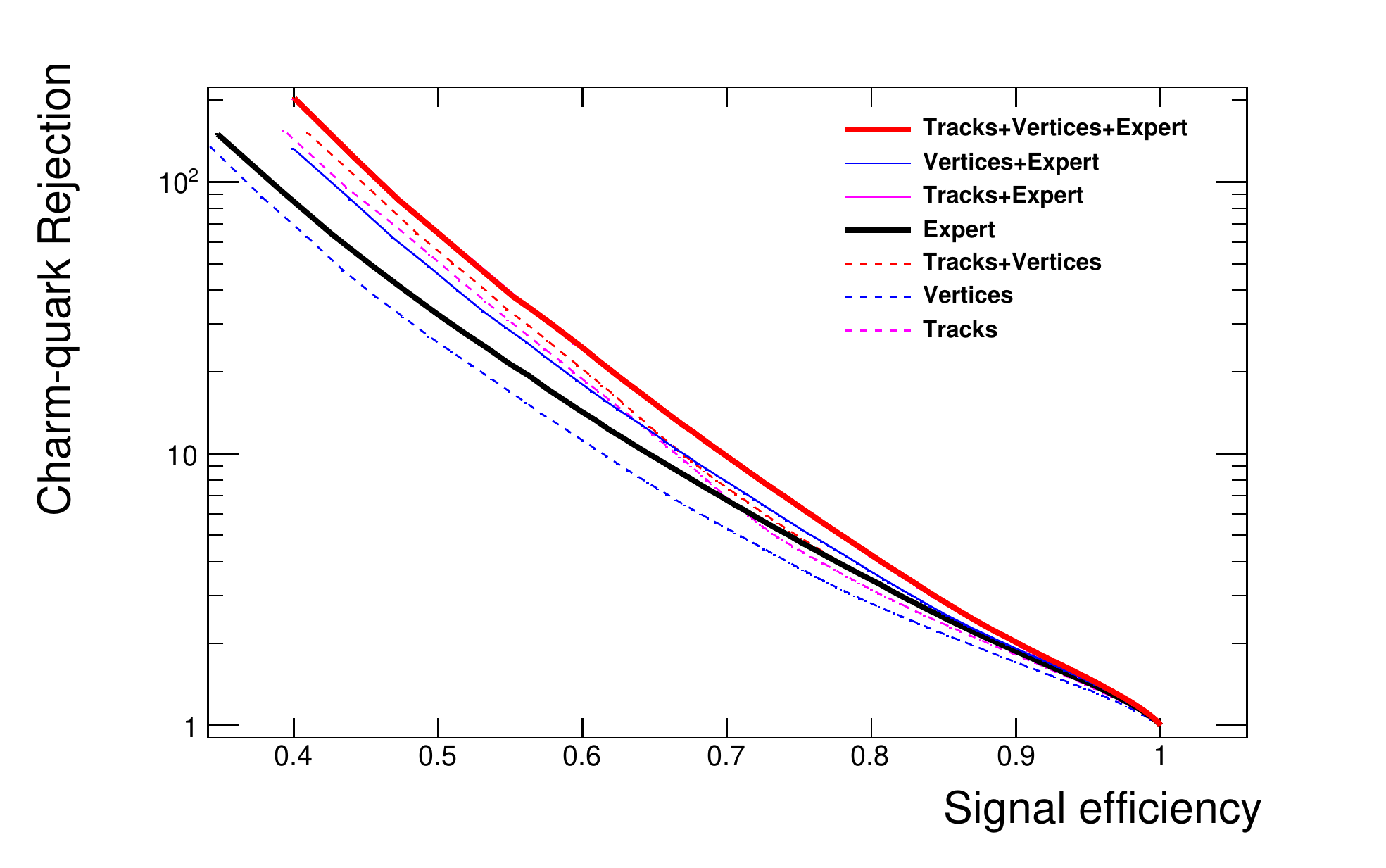}
\end{center}
\caption{Signal efficiency versus background rejection (inverse of efficiency) for deep networks trained on track-level, vertex-level or expert-level features. The top pane shows the performance for $b$-quarks versus light-flavor quarks, the bottom pane for $b$-quarks versus $c$-quarks.}
\label{fig:roc}
\end{figure*}

\begin{figure}
\begin{center}
\includegraphics[width=0.9\linewidth]{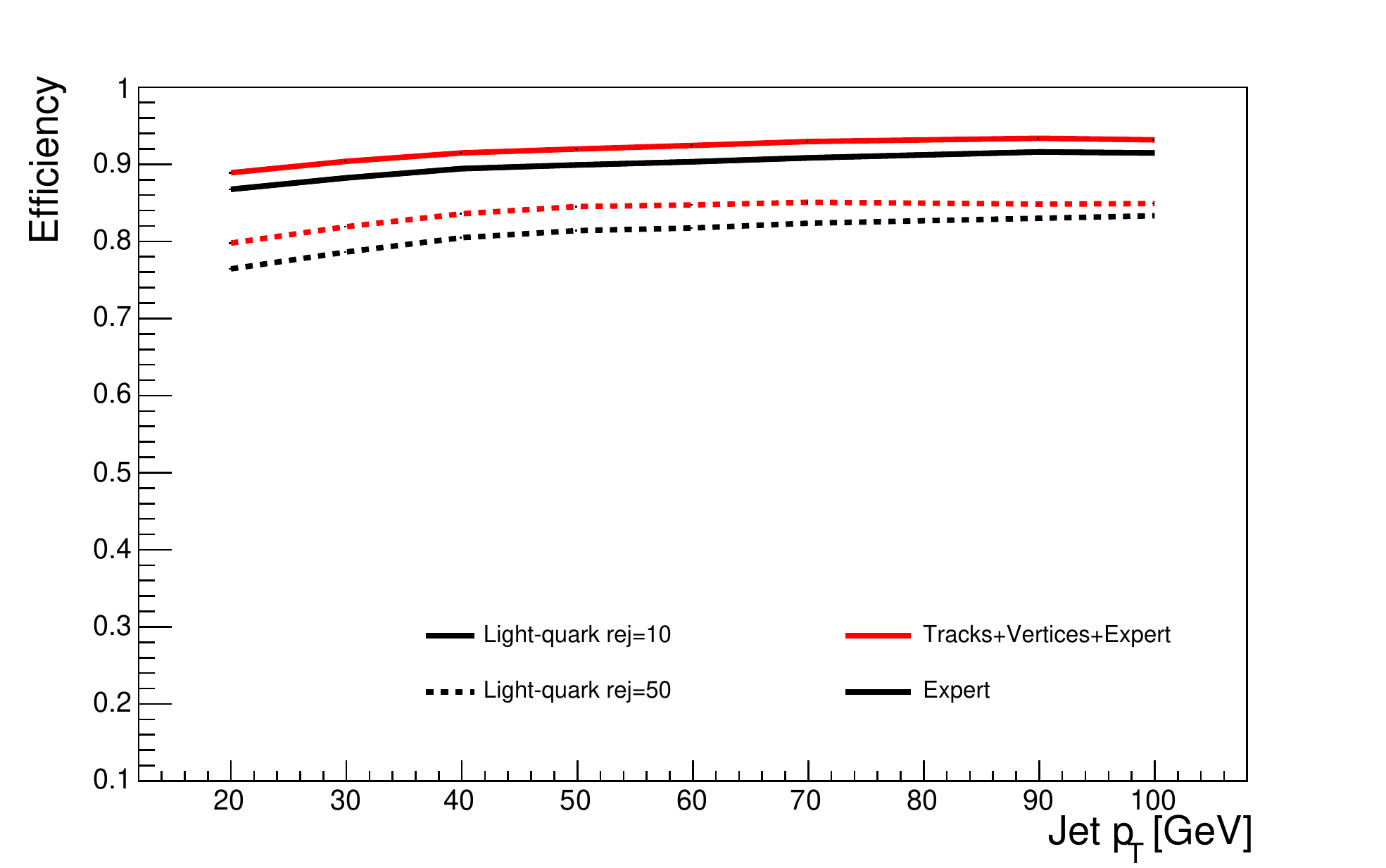}
\includegraphics[width=0.9\linewidth]{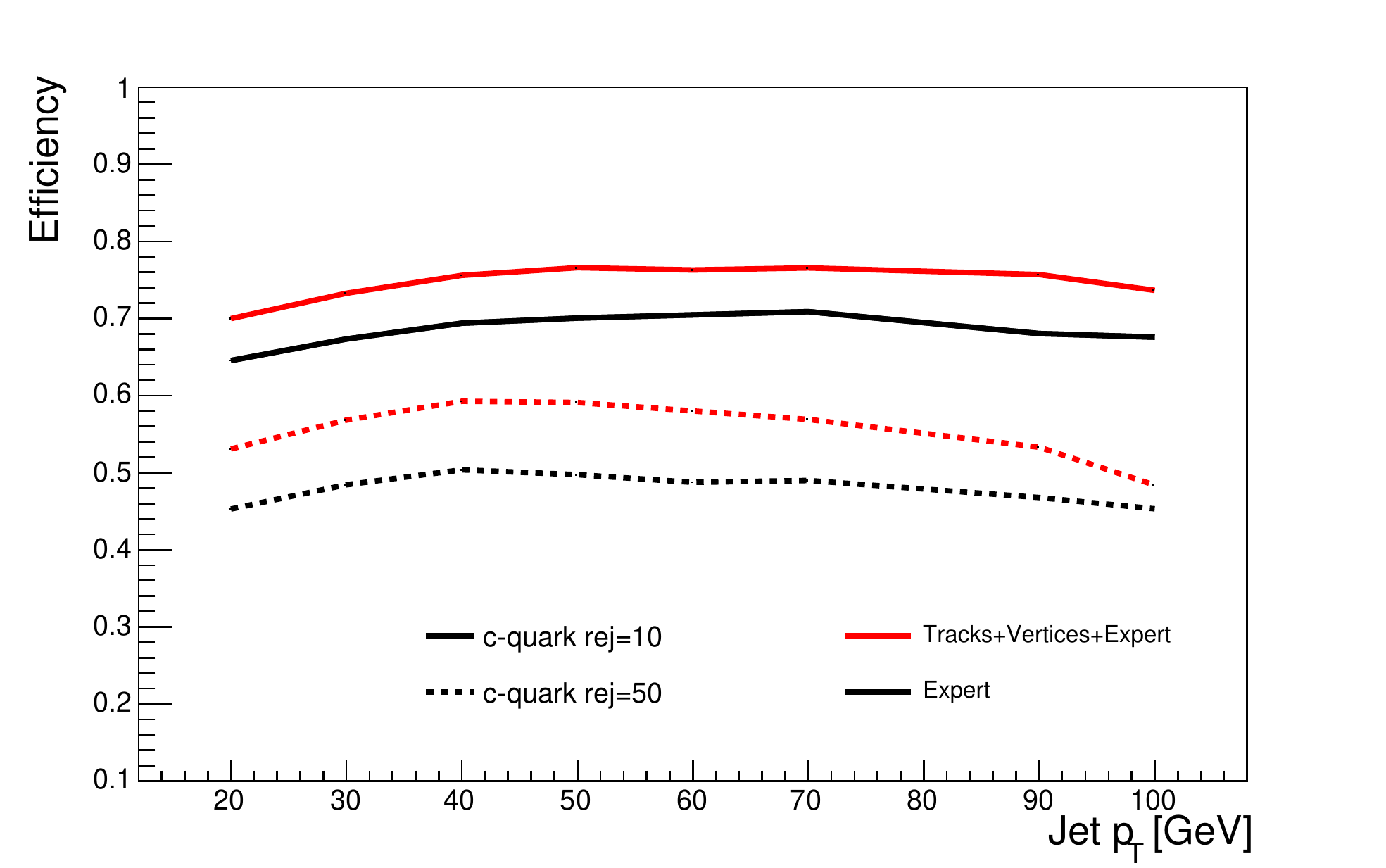}
\end{center}
\caption{Signal efficiency versus minimum jet $p_{\textrm{T}}$ relative to light quarks (top) or charm quarks (bottom). In each case, efficiency is shown for fixed values of background rejection for networks trained with only expert features or networks trained with all features (tracks, vertices and expert features).}
\label{fig:eff_pt}
\end{figure}

\begin{figure}
\begin{center}
\includegraphics[width=0.9\linewidth]{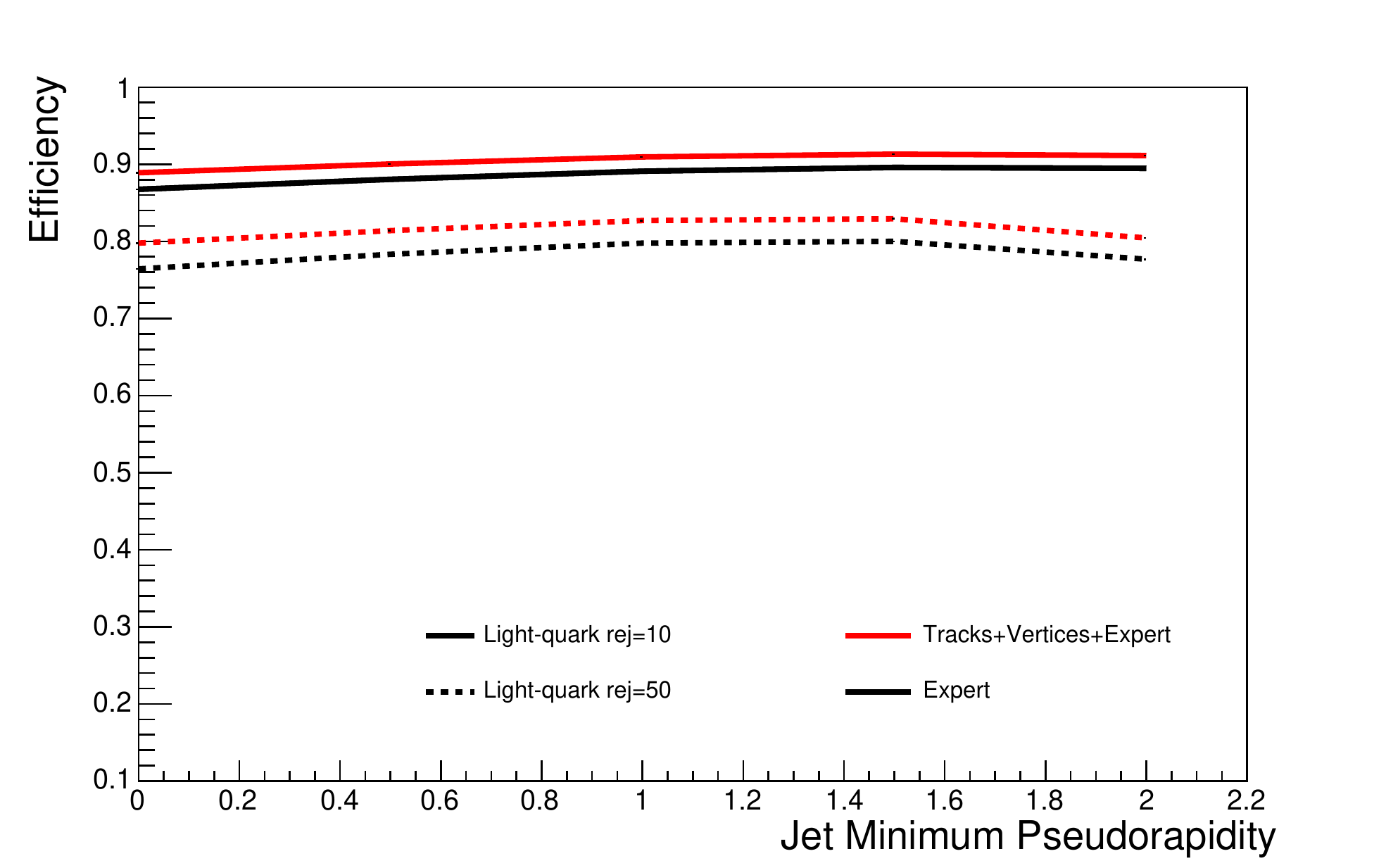}
\includegraphics[width=0.9\linewidth]{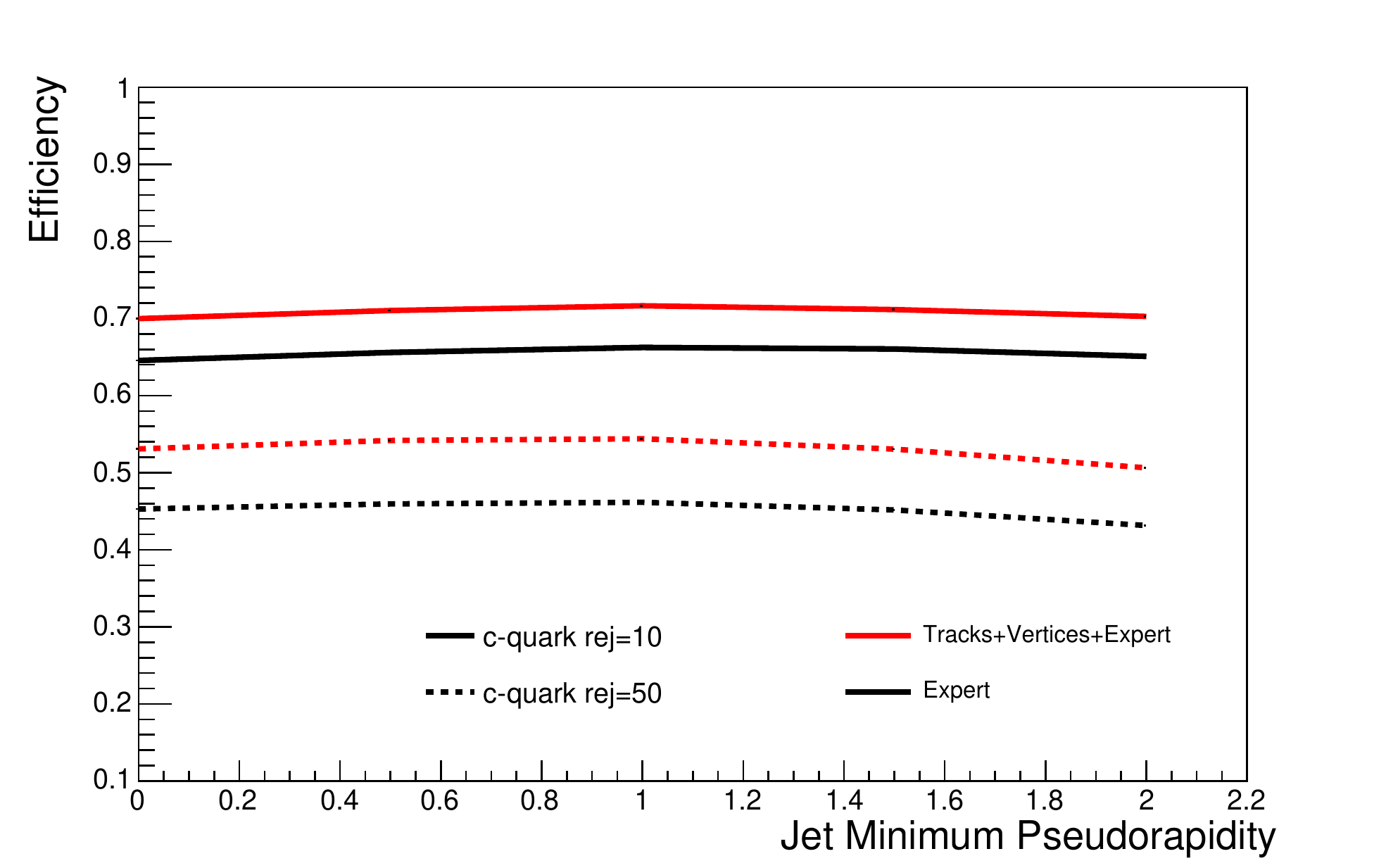}
\end{center}
\caption{Signal efficiency versus minimum jet pseudo-rapidity relative to light quarks (top) or charm quarks (bottom). In each case, efficiency is shown for fixed values of background rejection for networks trained with only expert features or networks trained with all features (tracks, vertices and expert features).}
\label{fig:eff_eta}
\end{figure}

\begin{figure}
\begin{center}
\includegraphics[width=0.9\linewidth]{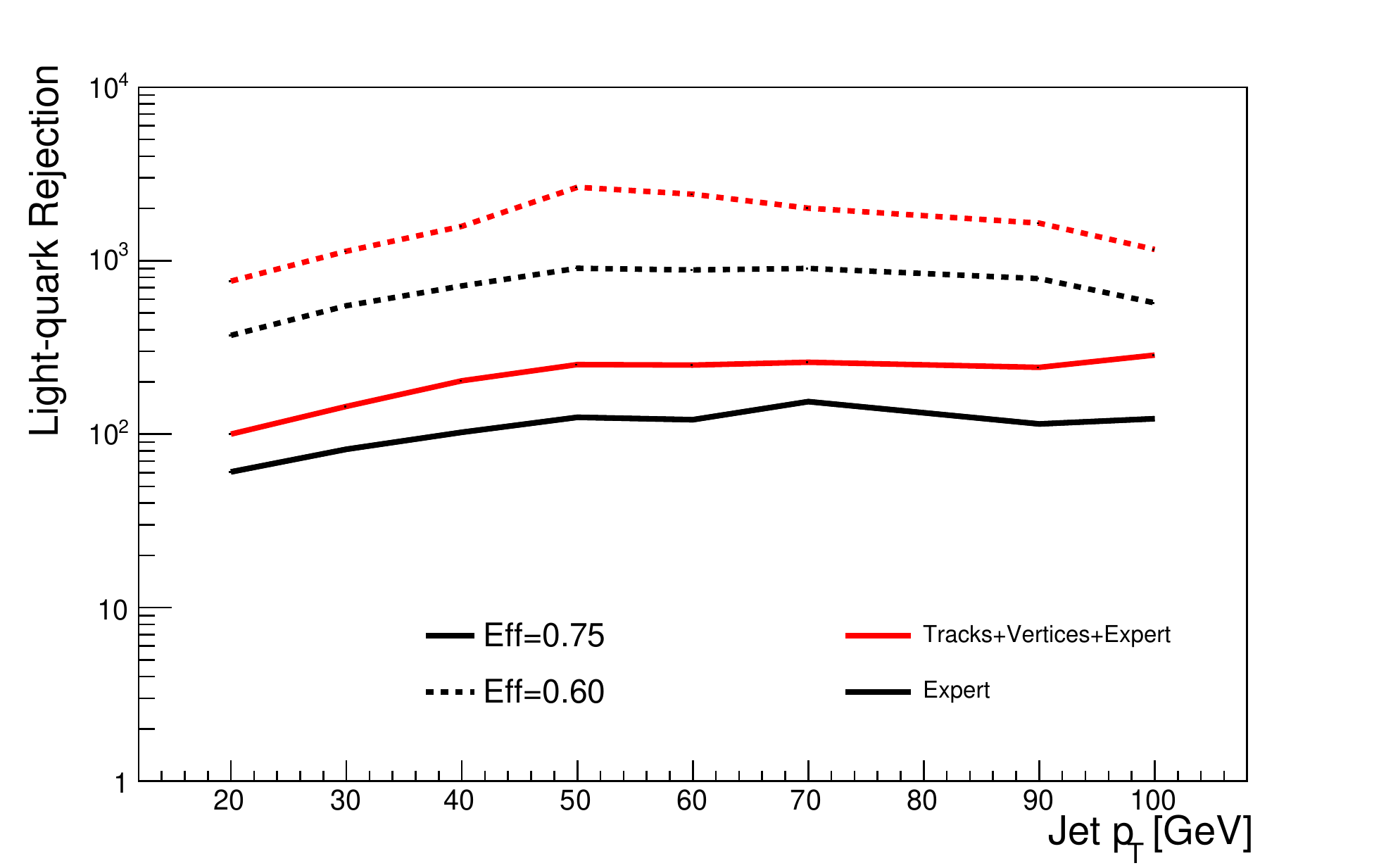}
\includegraphics[width=0.9\linewidth]{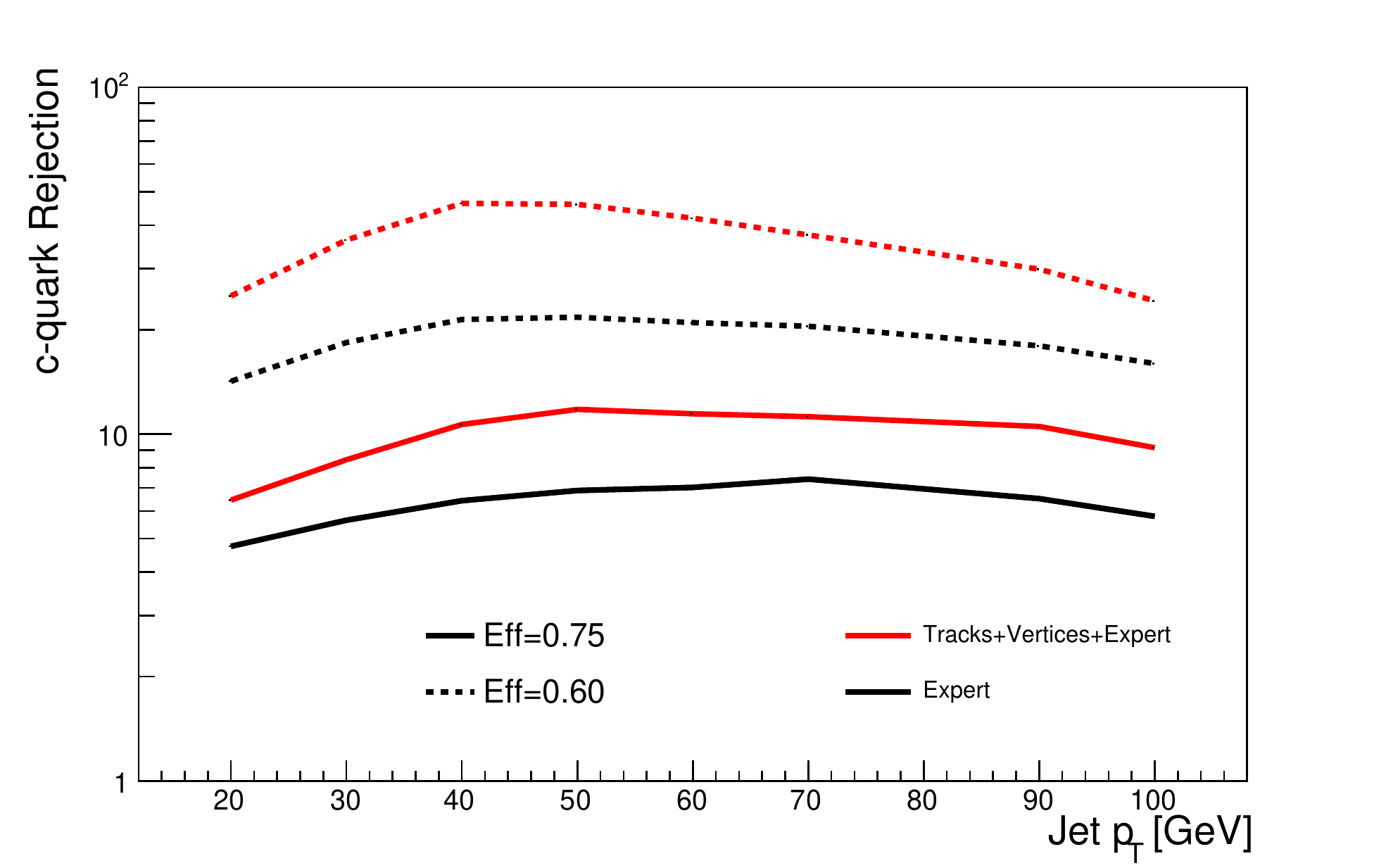}
\end{center}
\caption{Rejection of light quarks (top) or charm quarks (bottom) versus minimum jet $p_{\textrm{T}}$. In each case, rejection is shown for fixed values of signal efficiency for networks trained with only expert features or networks trained with all features (tracks, vertices and expert features).}
\label{fig:rej_pt}
\end{figure}

\begin{figure}
\begin{center}
\includegraphics[width=0.9\linewidth]{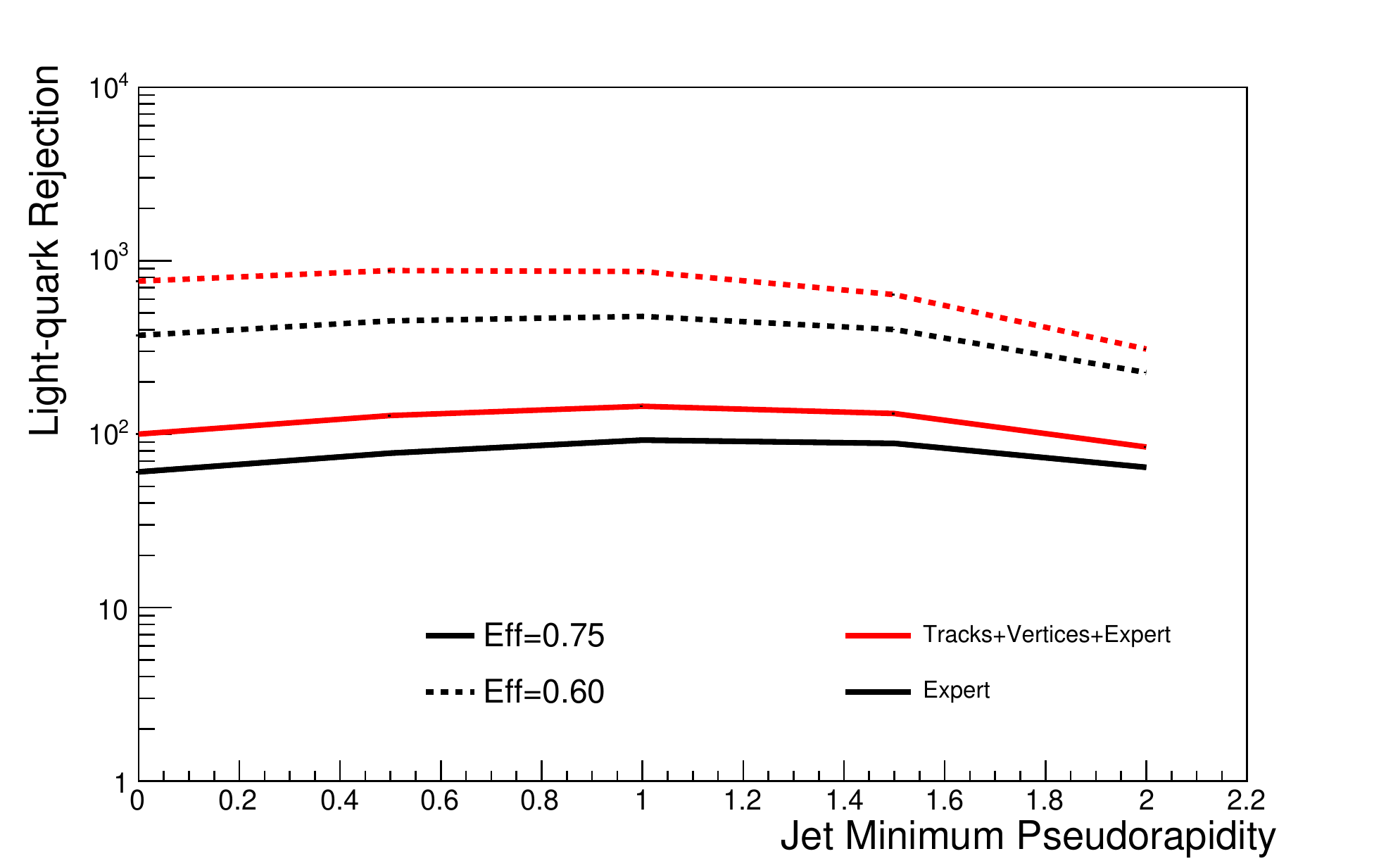}
\includegraphics[width=0.9\linewidth]{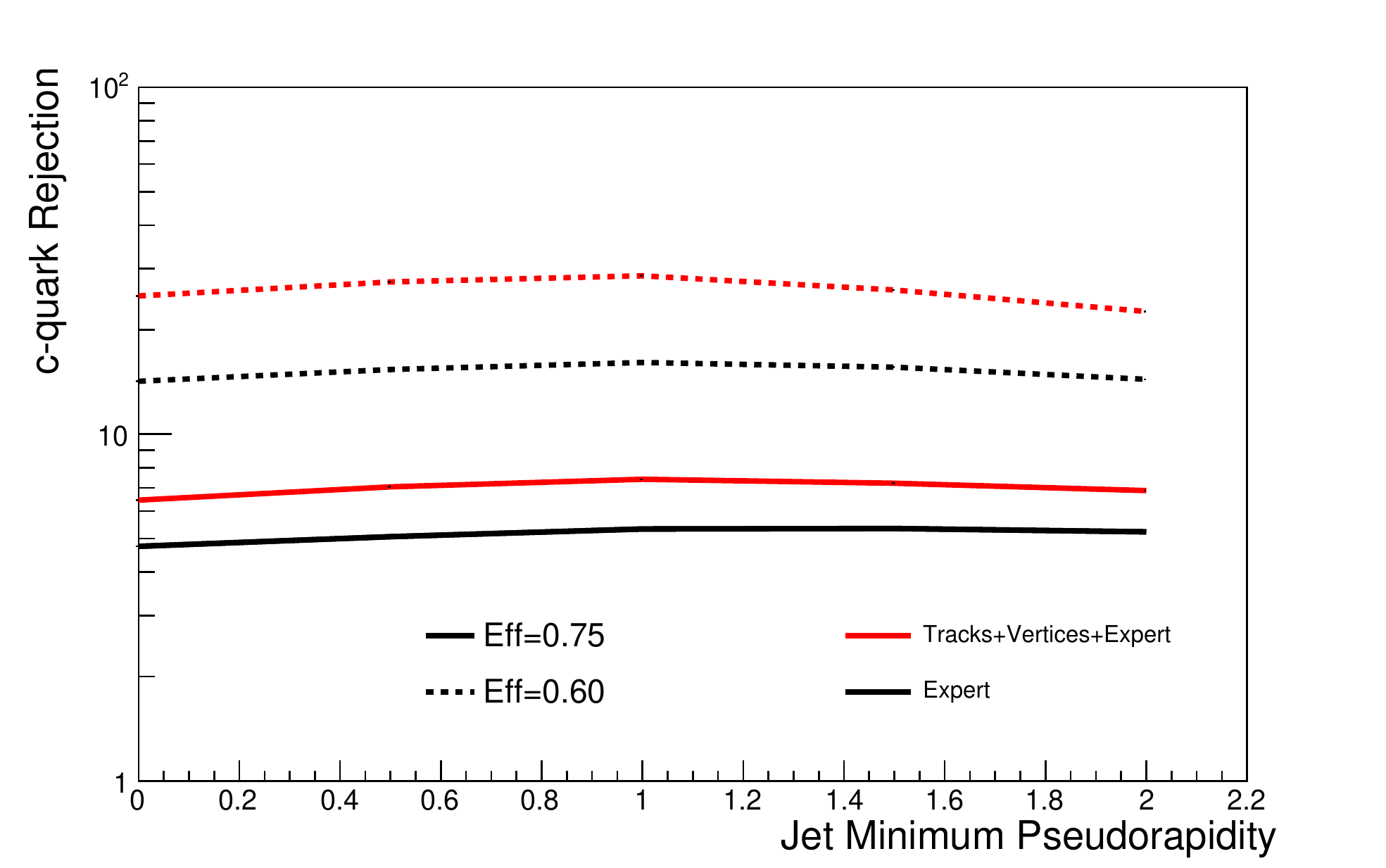}
\end{center}
\caption{Rejection of light quarks (top) or charm quarks (bottom) versus minimum jet pseudo-rapidity. In each case, rejection is shown for fixed values of signal efficiency for networks trained with only expert features or networks trained with all features (tracks, vertices and expert features).}
\label{fig:rej_eta}
\end{figure}

Final results are shown in Table~\ref{tab:performance}. The metric used is the Area Under the Curve (AUC), calculated in signal efficiency versus background efficiency, where a larger AUC indicates better performance.  In Fig.~\ref{fig:roc}, the signal efficiency is shown versus background rejection, the inverse of background efficiency.  Figures~\ref{fig:eff_pt} and~\ref{fig:eff_eta} show the efficiency versus jet $p_{\textrm{T}}$ and pseudorapidity for fixed values of background rejection. Figures~\ref{fig:rej_pt} and~\ref{fig:rej_eta} show the rejection versus jet $p_{\textrm{T}}$ and pseudorapidity for fixed values of signal efficiency. 

The results can be analyzed to draw conclusions regarding the power of the learning algorithms to extract information at different levels of preprocessing, and to compare the three learning approaches.

The state-of-the-art performance is represented by the networks which use only the expert-level features.  Networks using only tracking or vertexing features do not match this performance, though networks using both tracking and vertexing do slightly exceed it.  In addition, networks which combine expert-level information with track and/or vertex information outperform the expert-only benchmark, in some cases by a significant margin.

For any given set of features, the feedforward deep networks most often give the best performance, though in some cases by a small margin over the LSTM approach. This may be somewhat unexpected since LSTMs were created to handle variable sized input data as is the case here. We must note, however, that unlike truly sequential data like speech or text there is no natural order in the data that we are working on.
The tracks have been ordered by absolute $d_0$ significance, which tends to cluster tracks belonging to the same vertex, but a sequential model with this ordering may not be superior to processing tracks in parallel, as in the connected DNN with tied weights.

\begin{table}
\caption{ Performance results for networks using track-level, vertex-level or expert-level information. In each case the jet $p_T$ and pseudorapidity are also used. Shown for each method is the Area Under the Curve (AUC), the integral of the background efficiency versus signal efficiency, which have a statistical uncertainty of 0.001 or less. Signal efficiency and background rejections are shown in Figs.~\ref{fig:roc}-\ref{fig:rej_eta}. }
\begin{center}
\begin{tabular}{ccclll}
\hline\hline
 \multicolumn{3}{c}{Inputs}& Technique & AUC \\
Tracks & Vertices & Expert & \\
\hline
$\checkmark$ && & Feedforward & 0.916\\
$\checkmark$ &&& LSTM & 0.917\\
$\checkmark$ &&& Outer & 0.915\\
\hline
&$\checkmark$ & & Feedforward & 0.912\\
&$\checkmark$ & & LSTM & 0.911\\
&$\checkmark$ & & Outer & 0.911\\
\hline
$\checkmark$ &$\checkmark$& & Feedforward & 0.929\\
$\checkmark$ &$\checkmark$& & LSTM & 0.929\\
$\checkmark$ &$\checkmark$& & Outer & 0.928\\
\hline
&&$\checkmark$ &  Feedforward & 0.924\\
&&$\checkmark$ &  LSTM & 0.925\\
&&$\checkmark$ &  Outer & 0.924\\
\hline
$\checkmark$&&$\checkmark$   & Feedforward & 0.937 \\
$\checkmark$&&$\checkmark$& LSTM & 0.937 \\
$\checkmark$&&$\checkmark$& Outer & 0.936 \\
\hline
&$\checkmark$&$\checkmark$  & Feedforward & 0.931 \\
&$\checkmark$&$\checkmark$  & LSTM & 0.930 \\
&$\checkmark$&$\checkmark$  & Outer & 0.929 \\
\hline
$\checkmark$&$\checkmark$&$\checkmark$  & Feedforward & 0.939\\
$\checkmark$&$\checkmark$&$\checkmark$ & LSTM & 0.939 \\
$\checkmark$&$\checkmark$&$\checkmark$ & Outer & 0.937 \\
\hline\hline
\end{tabular}
\end{center}
\label{tab:performance}
\end{table}

\begin{figure*}
\begin{center}
\includegraphics[width=0.2\linewidth]{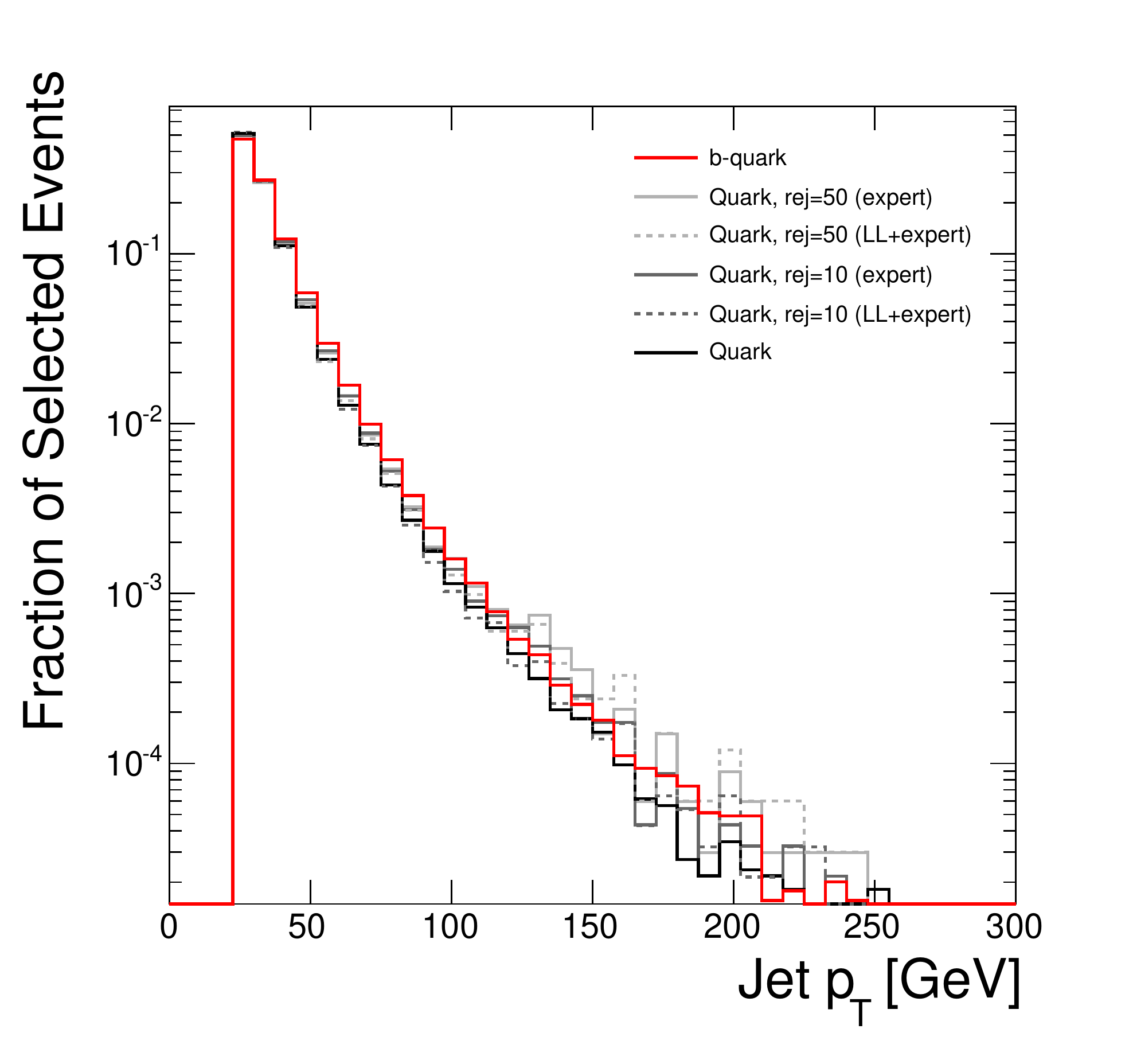}
\includegraphics[width=0.2\linewidth]{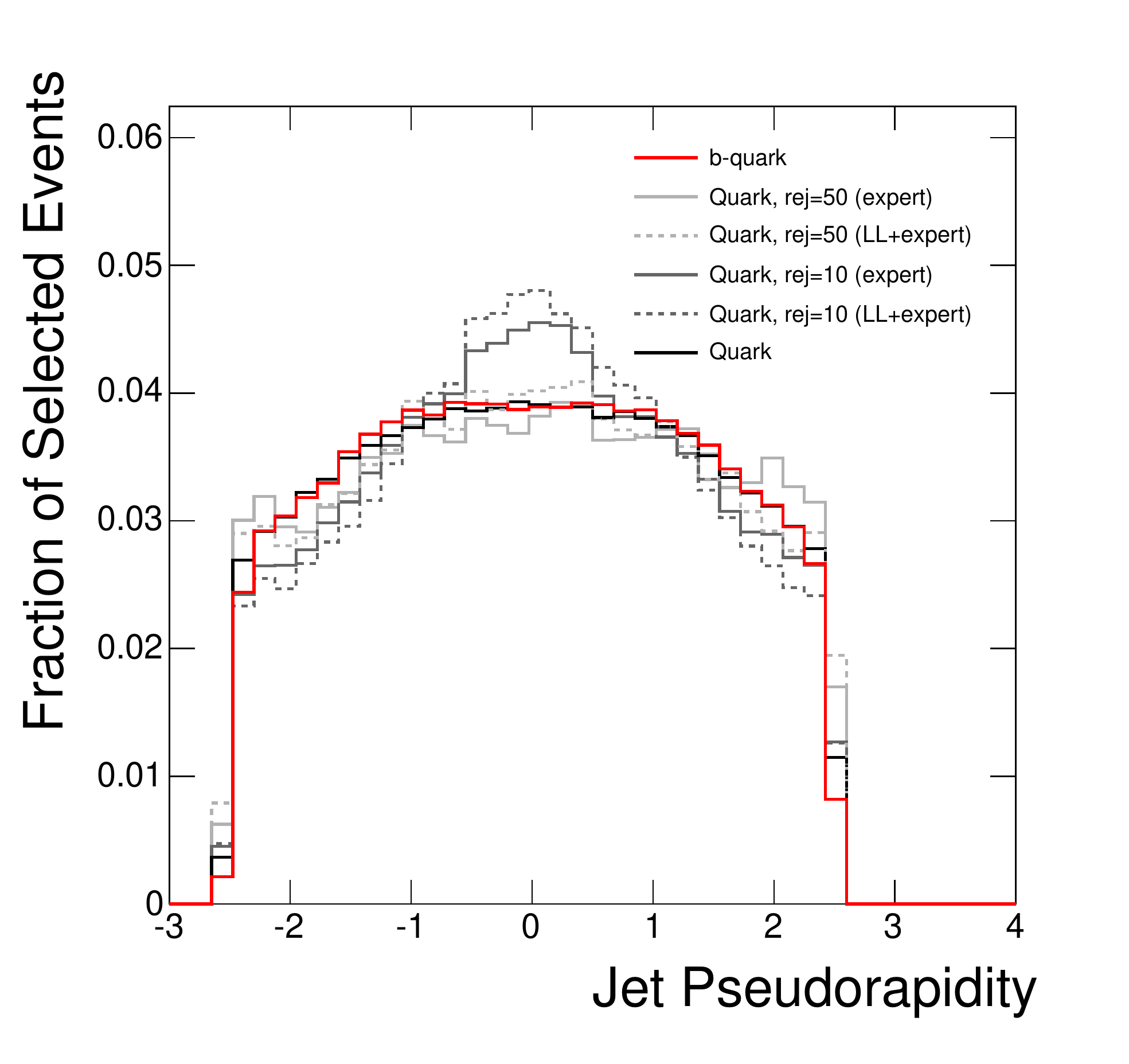}
\includegraphics[width=0.2\linewidth]{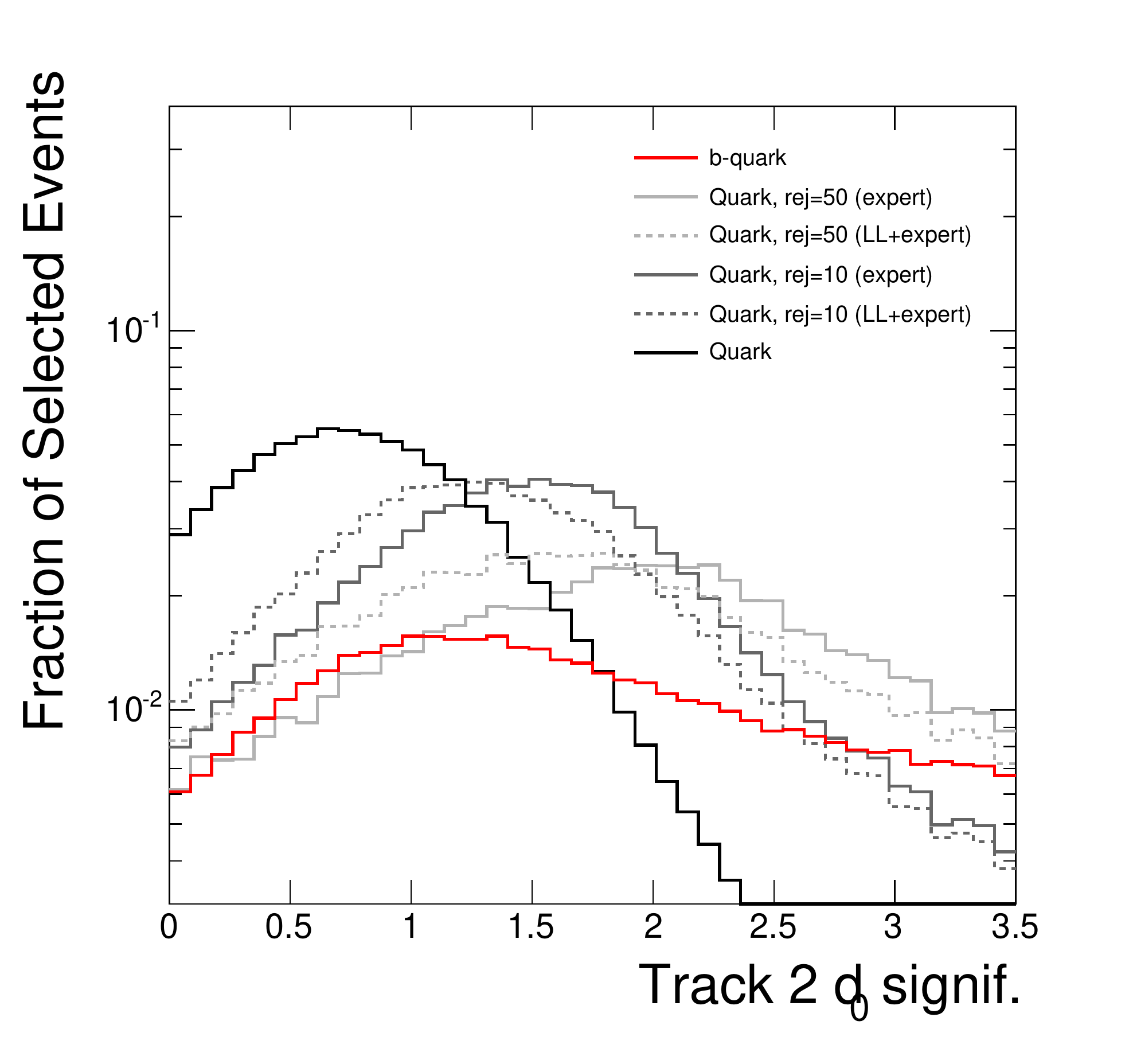}
\includegraphics[width=0.2\linewidth]{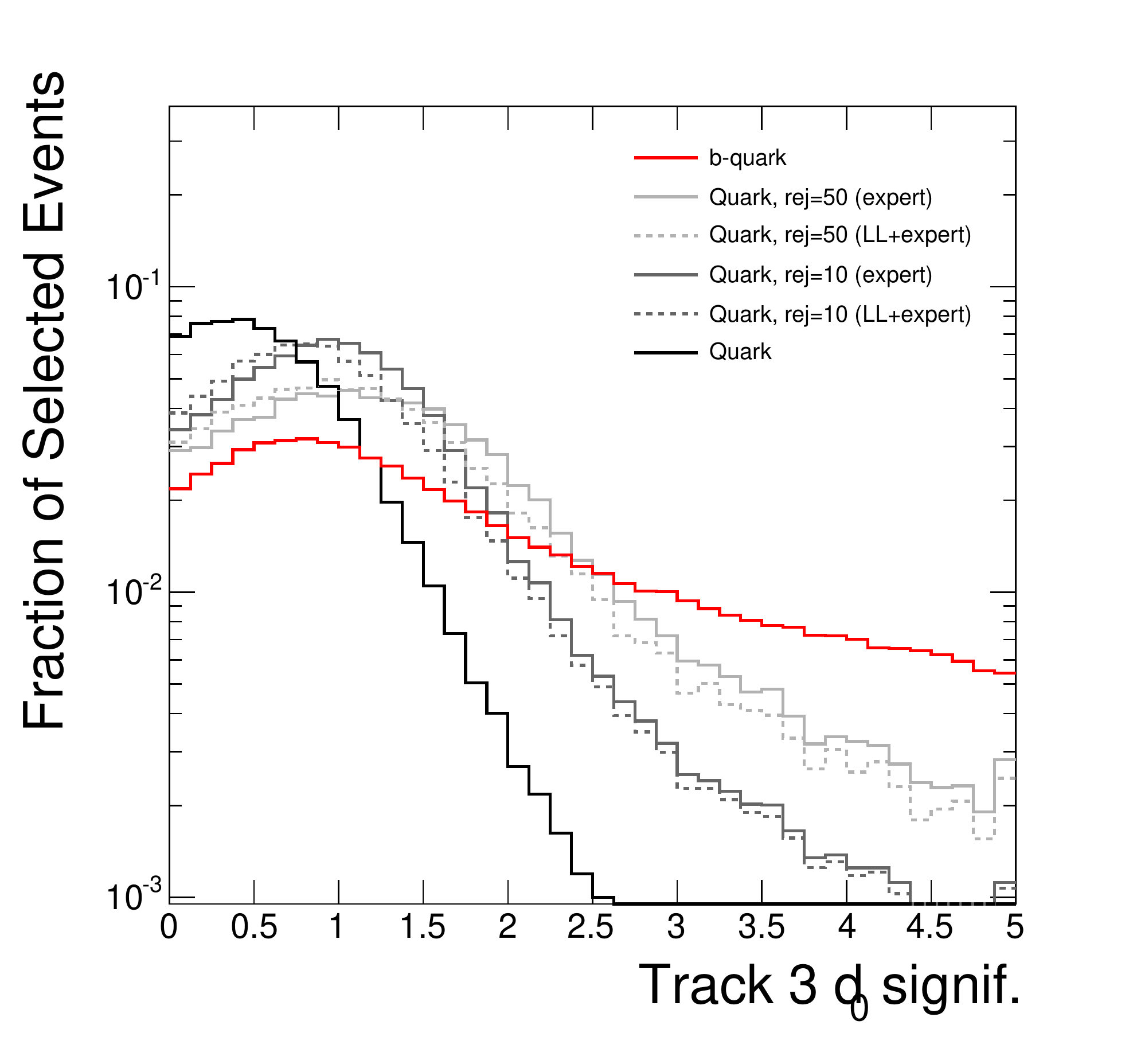}\\
\includegraphics[width=0.2\linewidth]{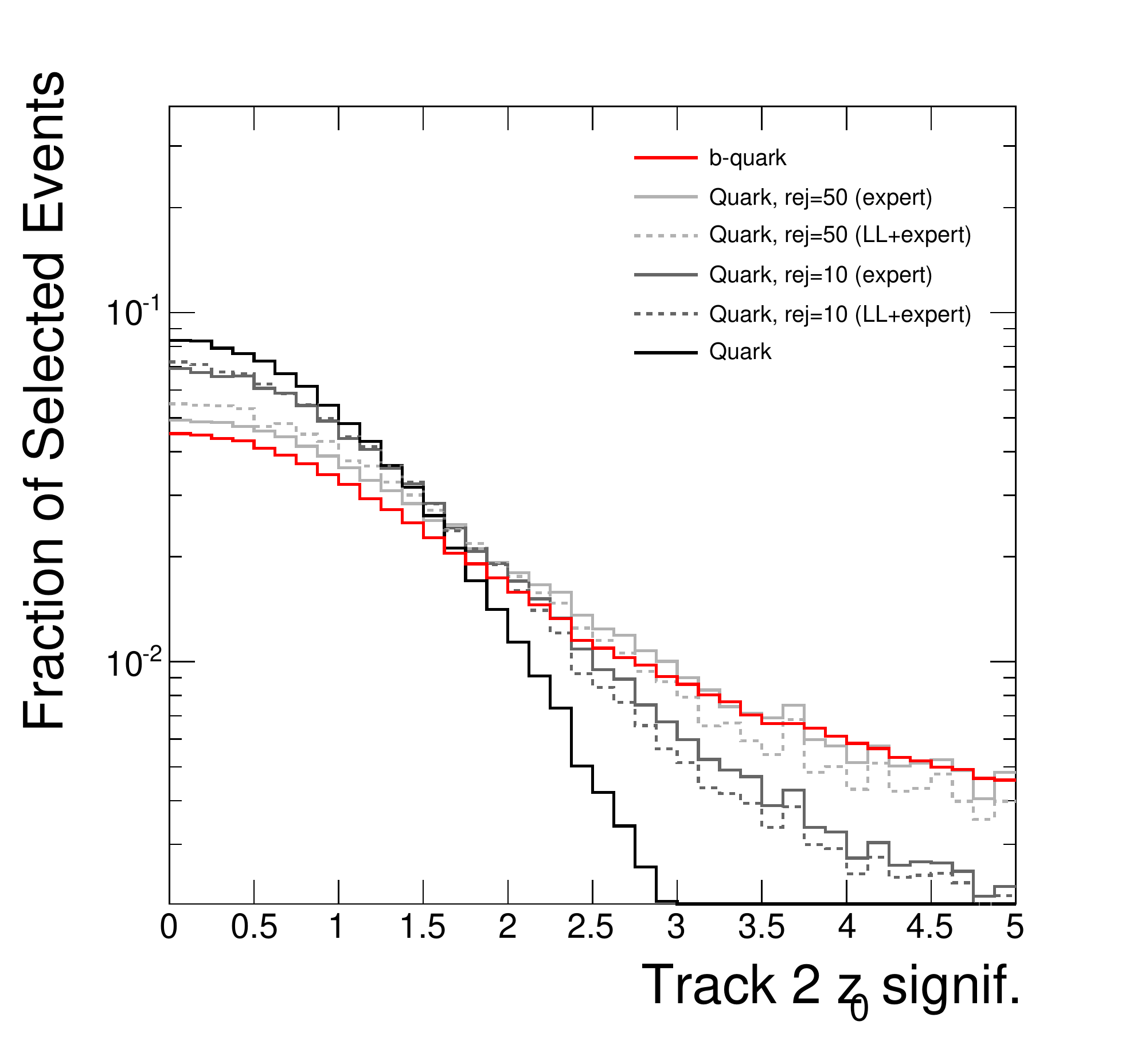}
\includegraphics[width=0.2\linewidth]{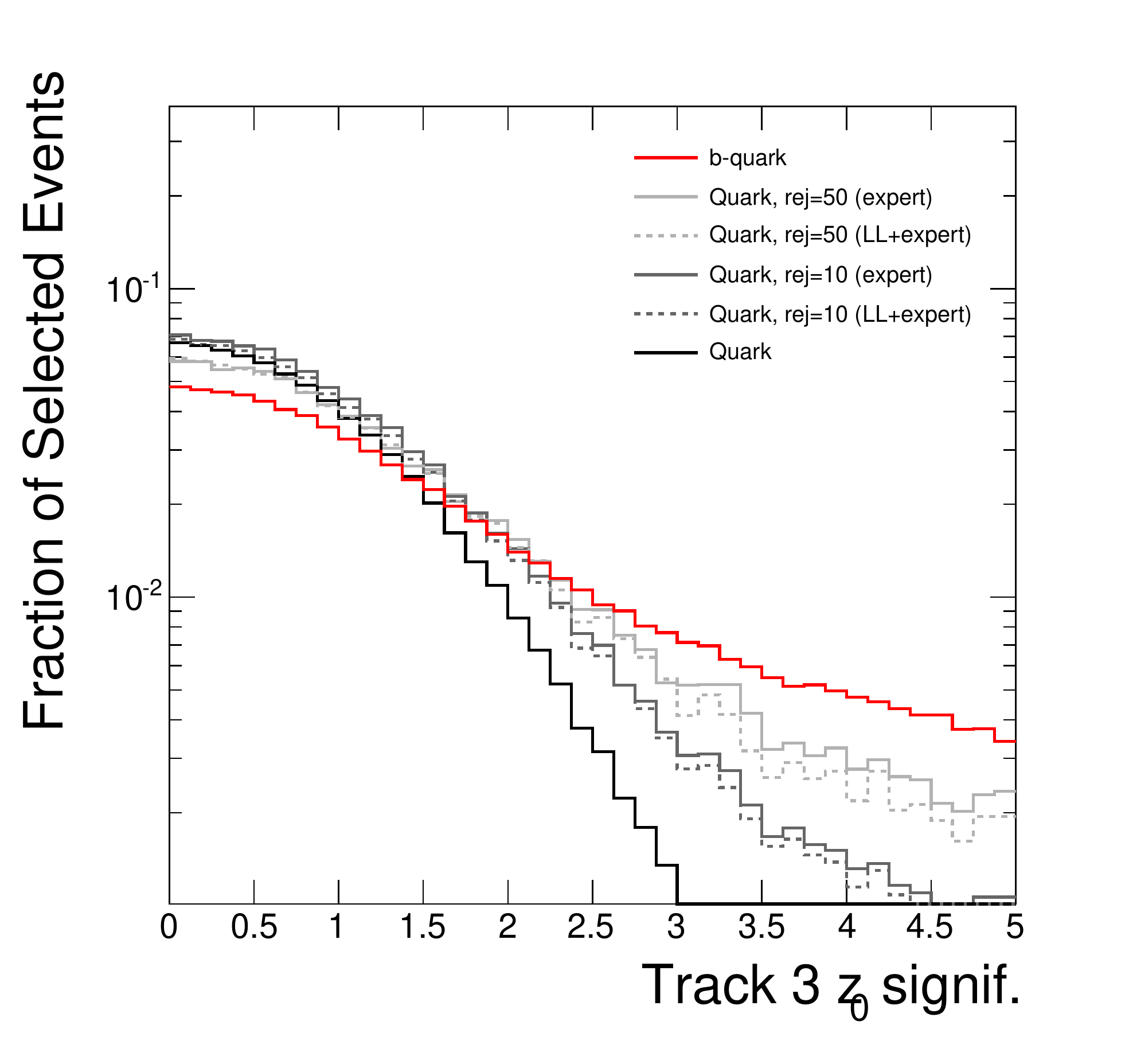}
\includegraphics[width=0.2\linewidth]{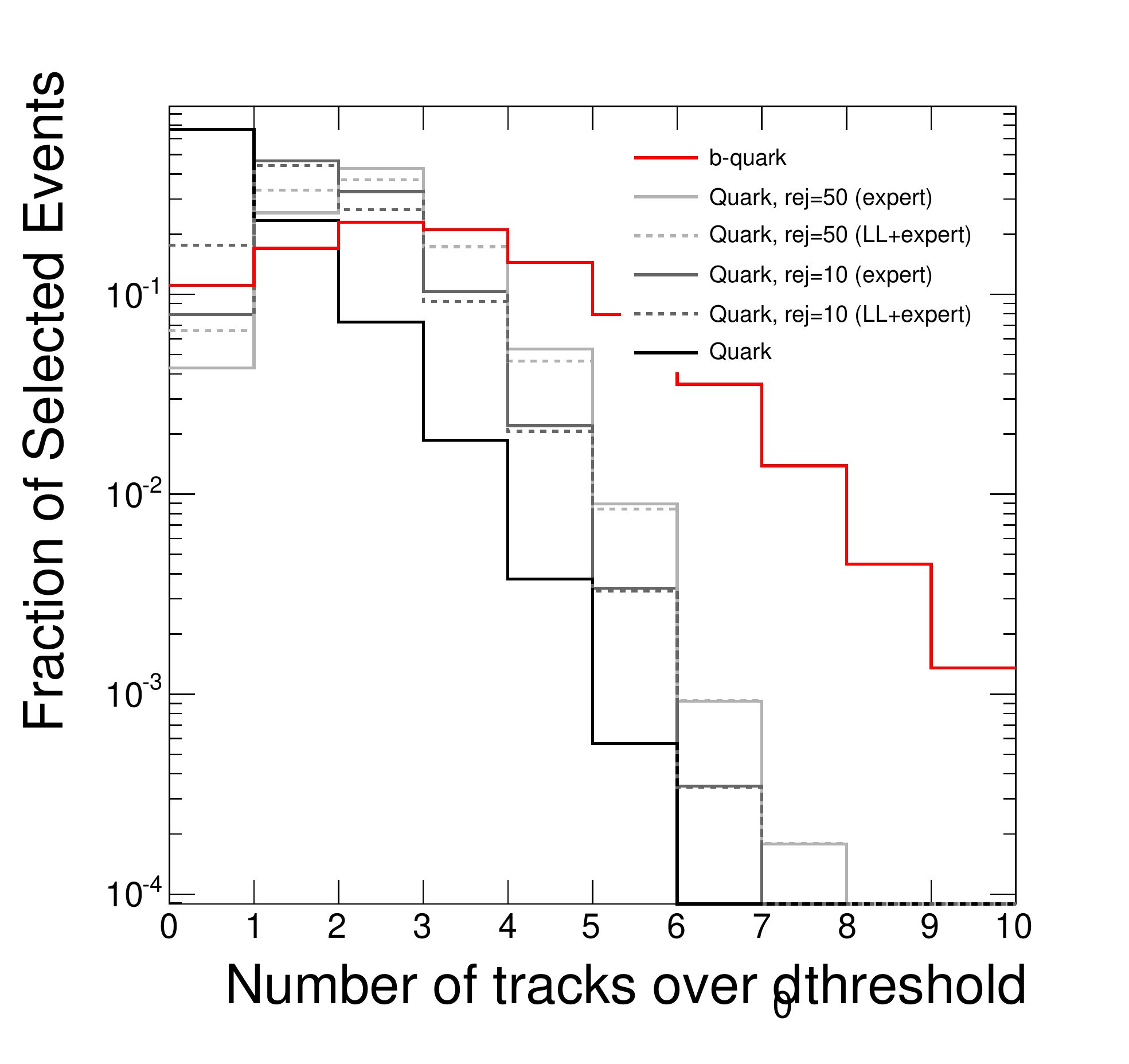}
\includegraphics[width=0.2\linewidth]{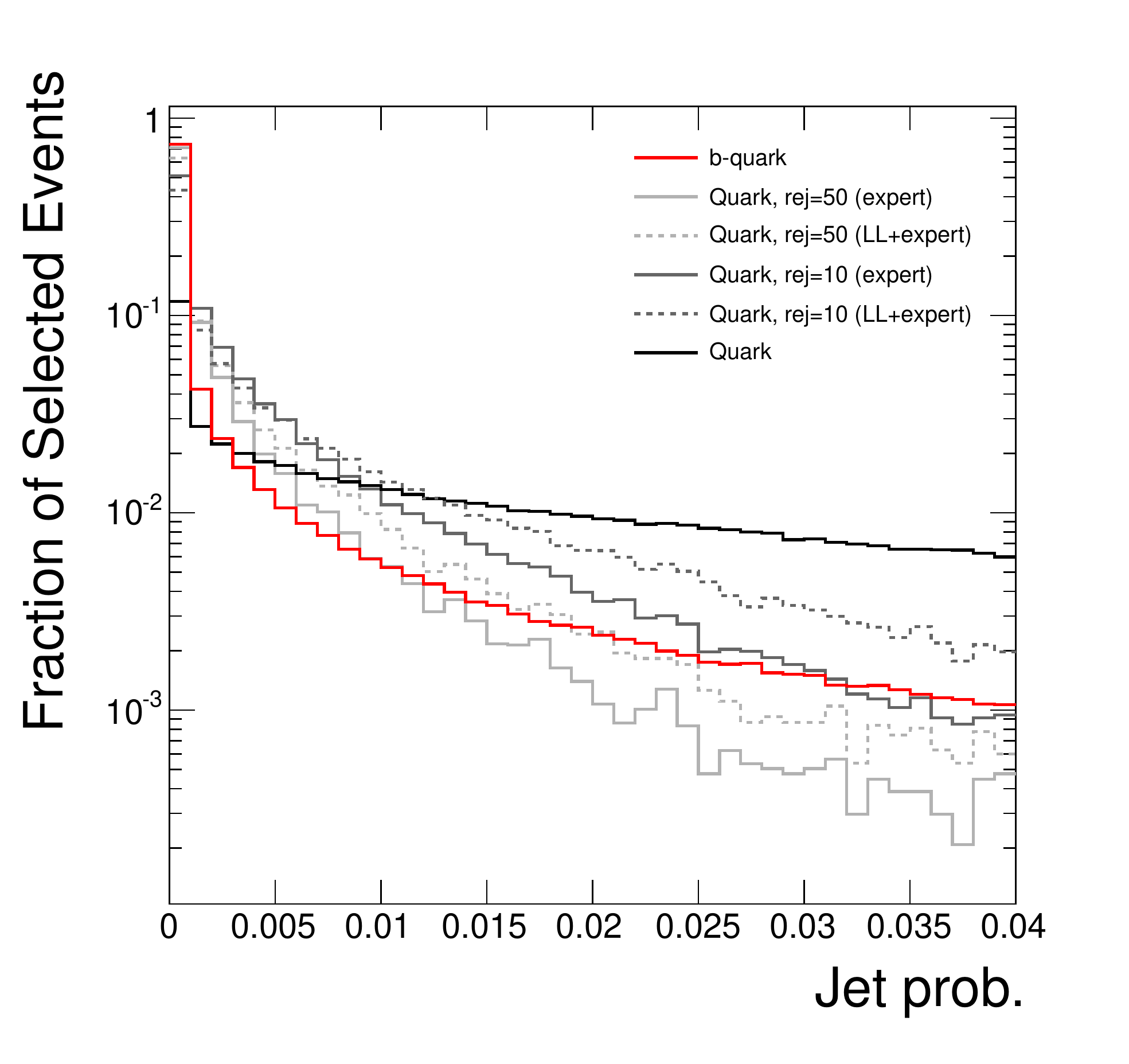}\\
\includegraphics[width=0.2\linewidth]{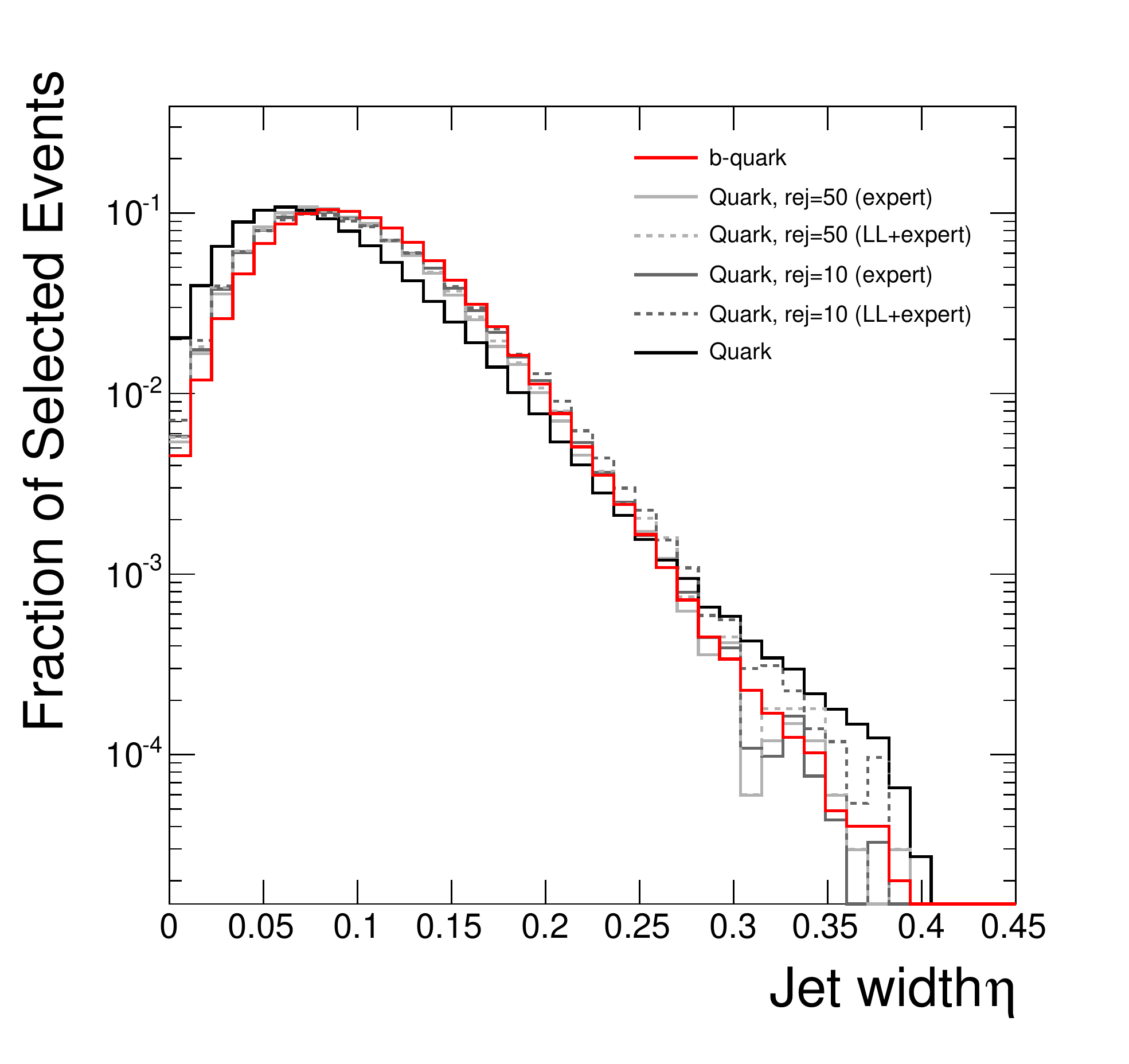}
\includegraphics[width=0.2\linewidth]{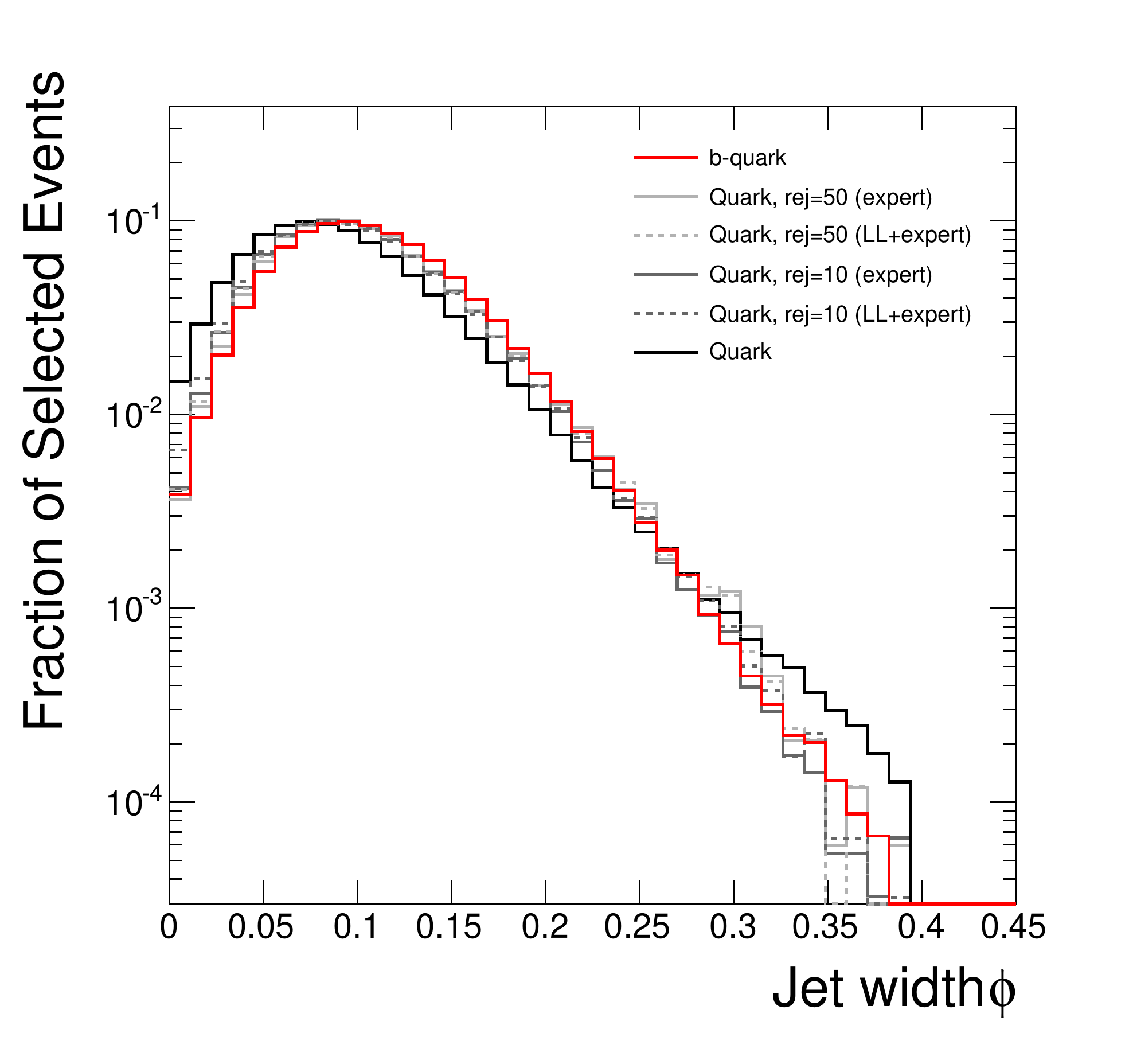}
\includegraphics[width=0.2\linewidth]{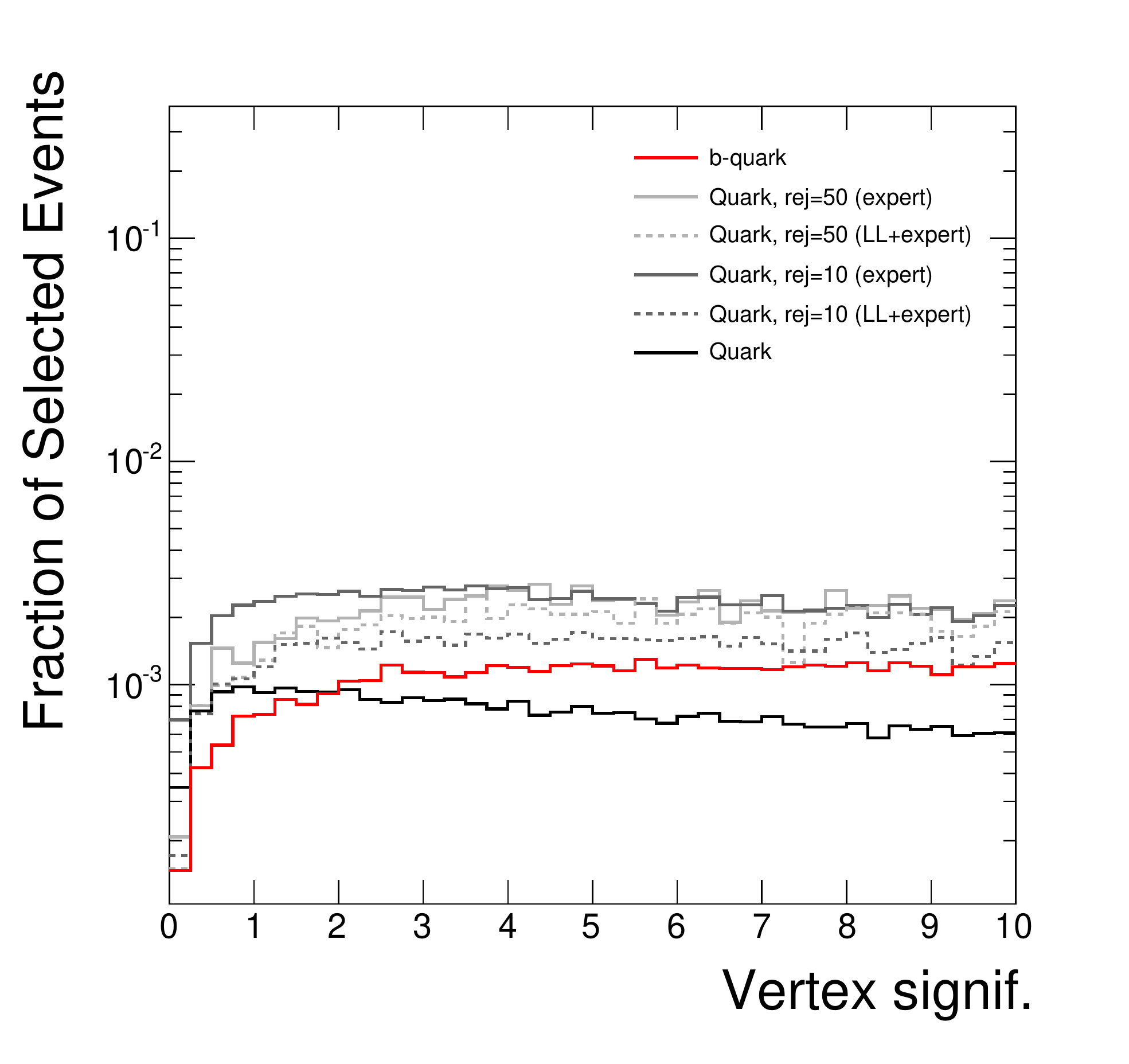}
\includegraphics[width=0.2\linewidth]{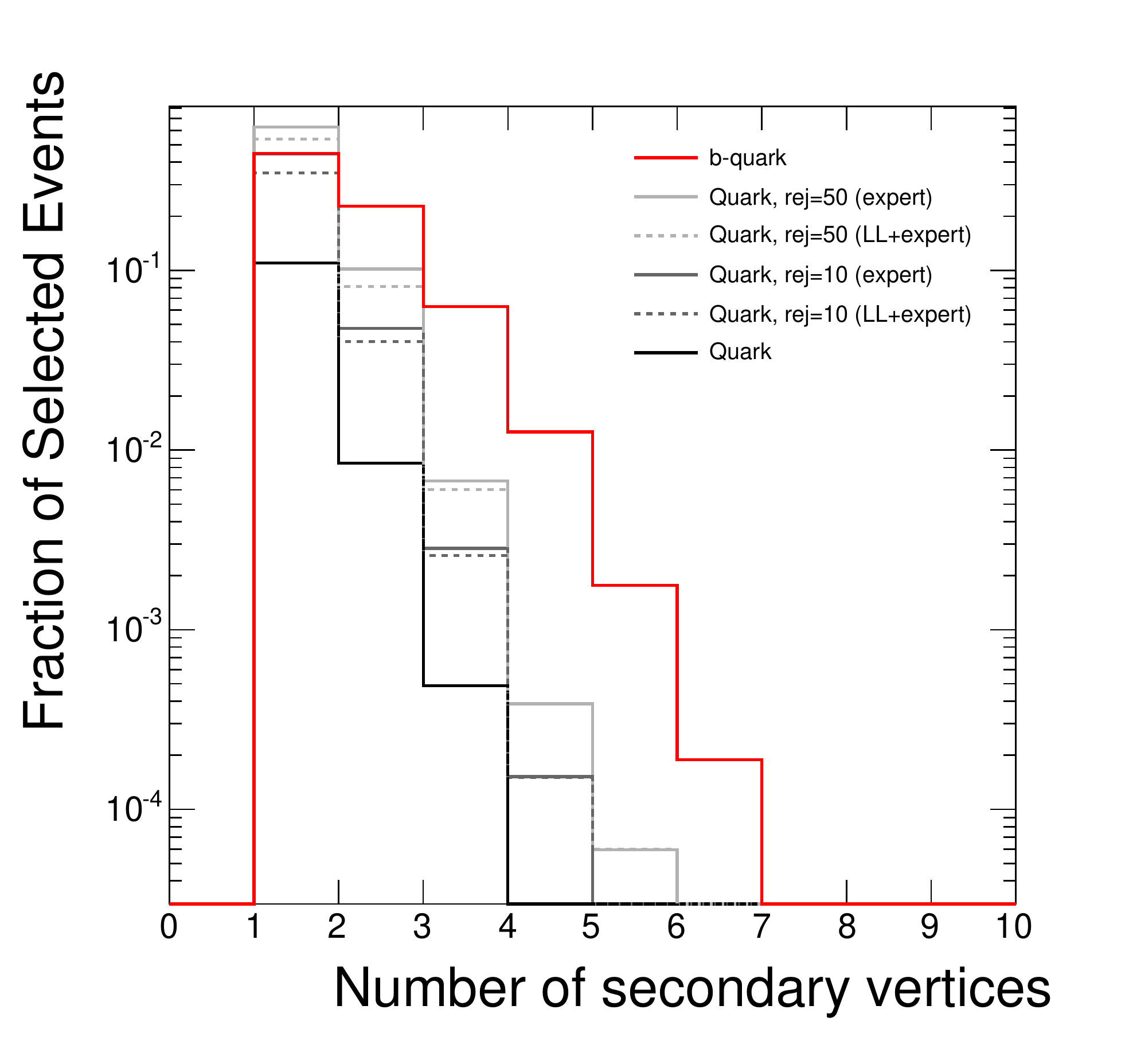}\\
\includegraphics[width=0.2\linewidth]{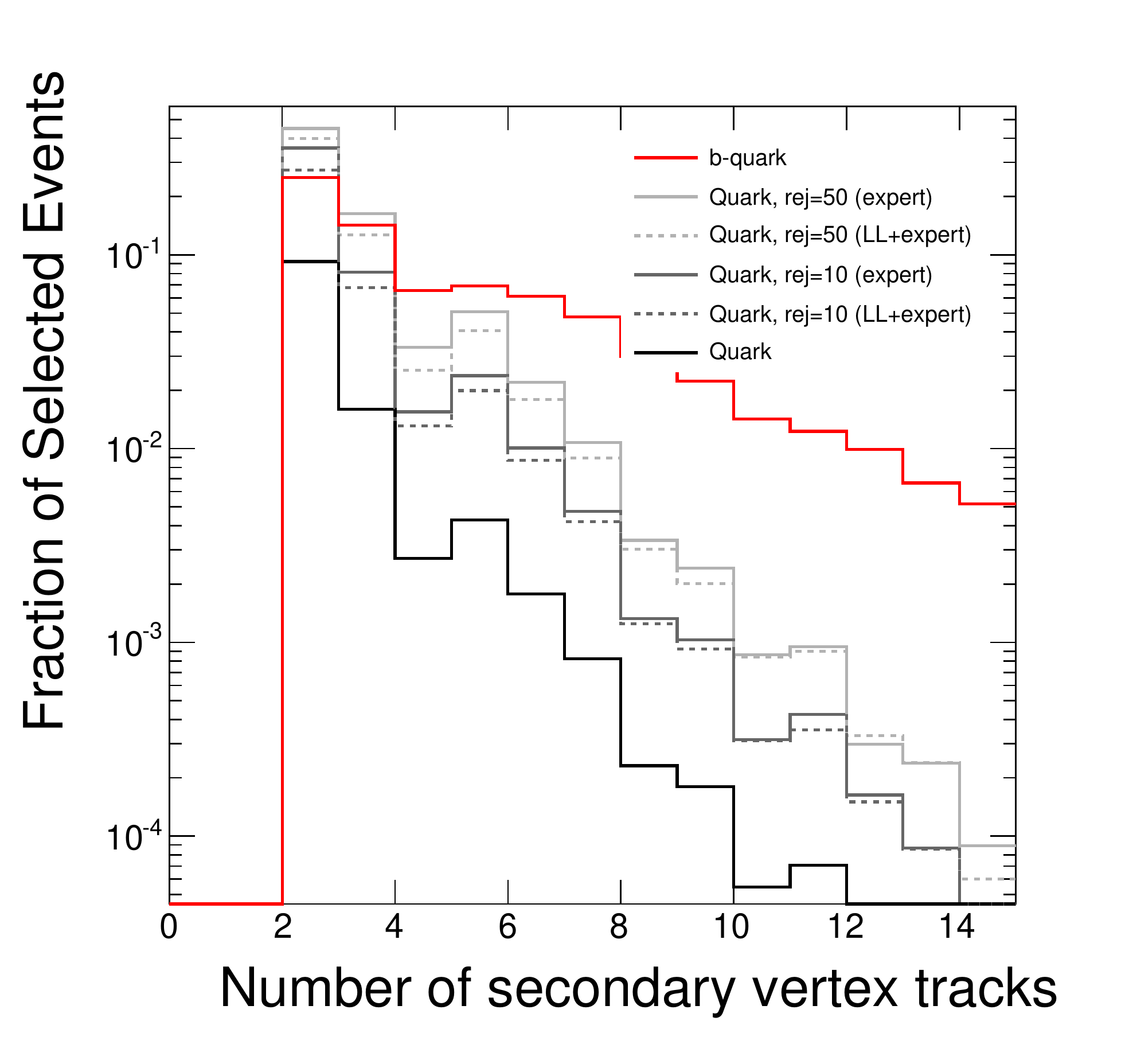}
\includegraphics[width=0.2\linewidth]{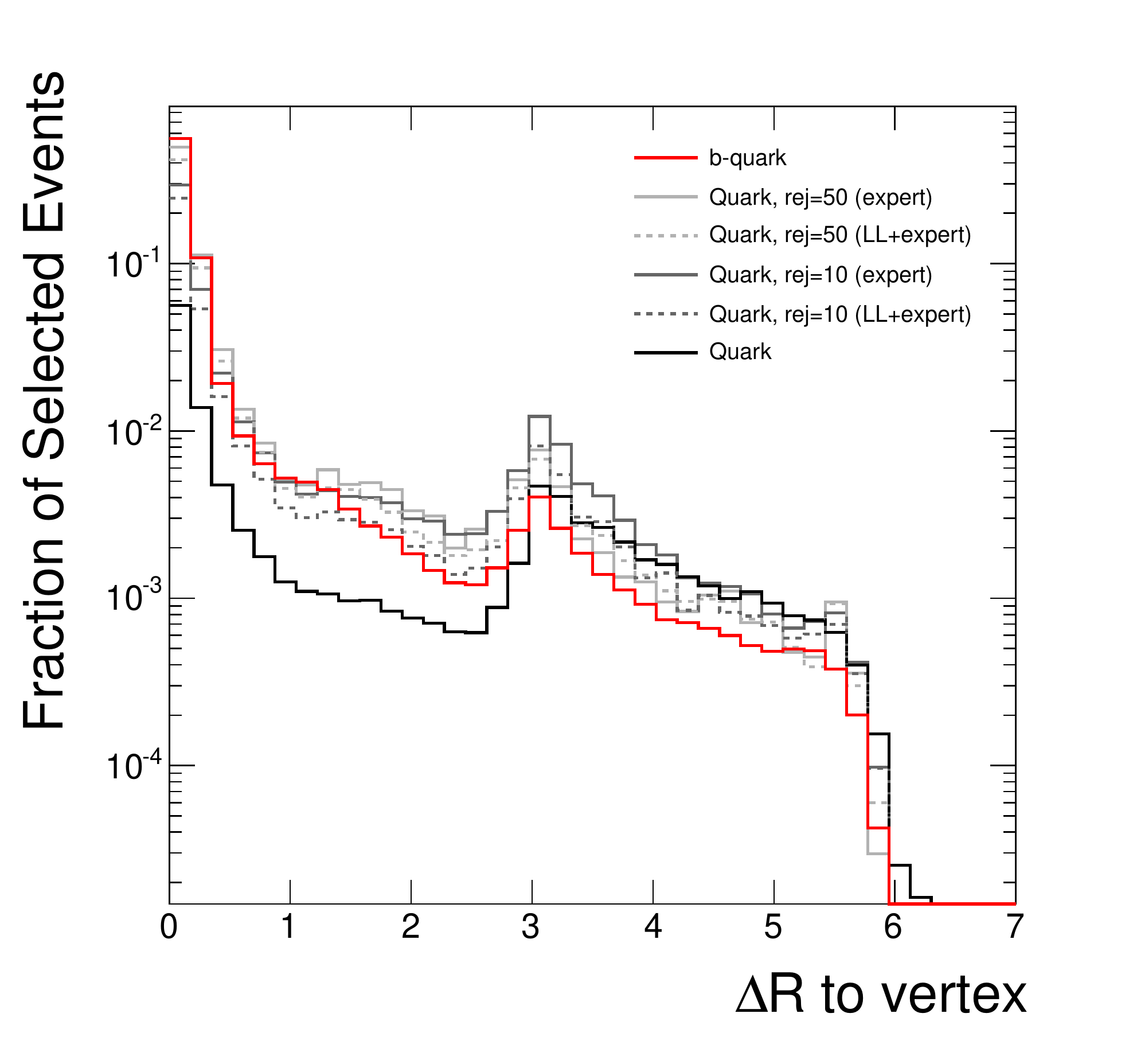}
\includegraphics[width=0.2\linewidth]{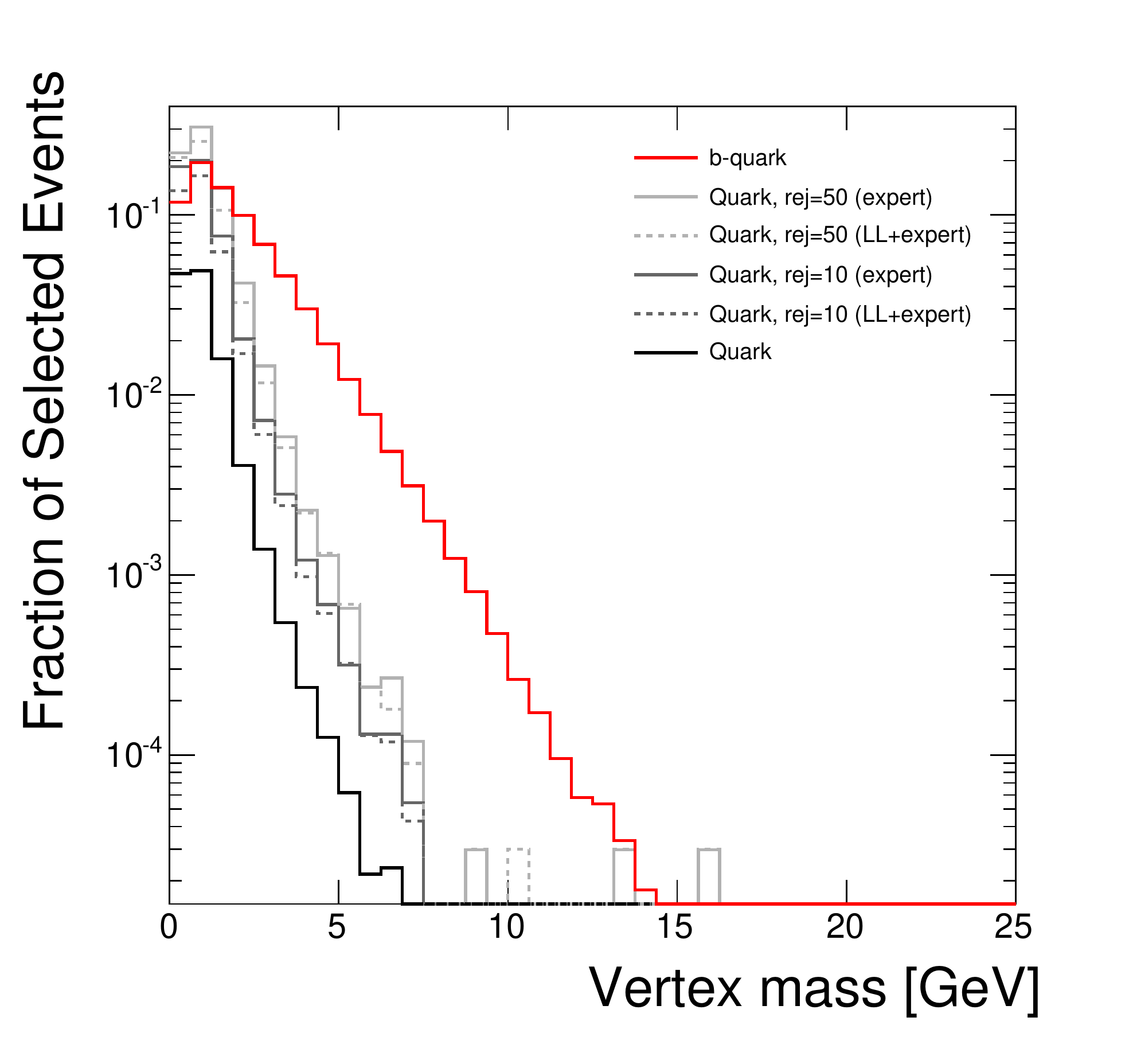}
\includegraphics[width=0.2\linewidth]{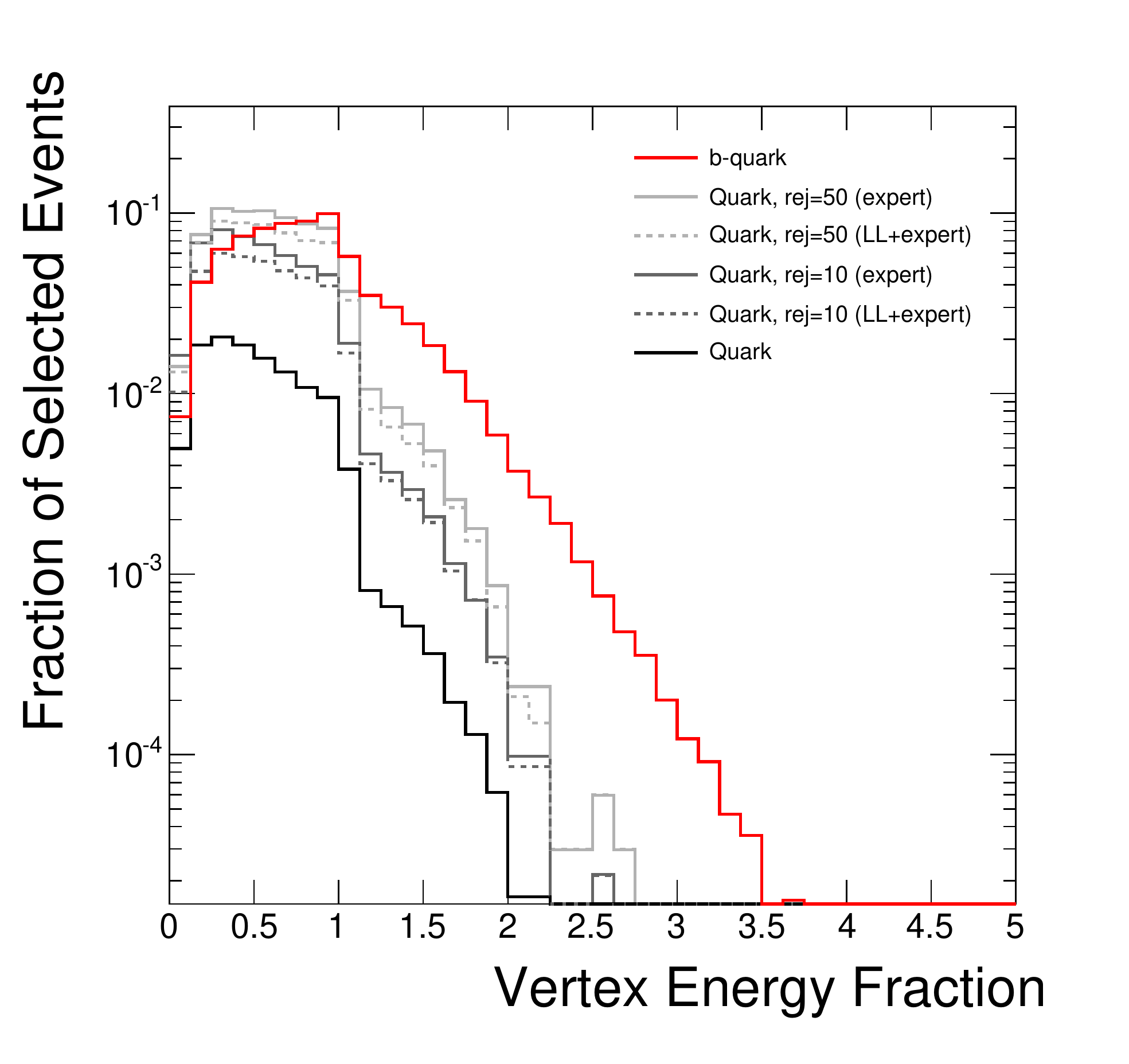}
\end{center}
\caption{Distributions of expert-level features for heavy-flavor and light-flavor classes. Also shown are distributions of light-flavor and charm jets surviving  network threshold selections chosen to given rejection of 10 and 50, for networks using only expert information and networks using expert information in addition to lower-level information.}
\label{fig:slice}
\end{figure*}

While one cannot probe the strategy of the ML algorithm, it is possible to compare distributions of events categorized as signal-like by the different algorithms in order to understand how the classification is being accomplished.  To compare distributions between different algorithms,  we study simulated events with equivalent background rejection, see Fig.~\ref{fig:slice} for a comparison of the selected regions in the expert features for classifiers with and without the lower-level information.

\section{Discussion}

Our experiments support four conclusions.

{\bf The existing expert strategies for dimensional reduction sacrifice or distort useful information.}  Networks which include lower-level information outperform networks using exclusively higher-level information.  For example, if the vertex-level information contained all of the classification power of the track-level information but with lower dimensionality, one would expect the vertex-only network to match the performance of the tracks-and-vertex network, as the lower-dimensional problem should be simpler to learn. Instead, networks using tracks and vertices outperform those which use only vertices.  Similarly, networks using tracks and expert features outperform those with only expert features. We note that these conclusions apply to the expert strategies considered here, and in the case of the simulated environment we have studied; however, we feel that both are representative of the current state-of-the-art.

{\bf The task remains a challenge for deep networks.} Networks which use only the lower-level information do not match the performance of networks which use the higher-level information.  Since the higher-level features are strict functions of the lower-level features, the lower-level features are a superset of the information contained in the high-level features. The performance of the networks which use the high-level features then provides a baseline against which to measure the ability of the network to extract the relevant information in the more difficult higher-dimensional space of lower-level features.  Networks using only track information do not match the performance of those which use only the high-level features (but note that track-only networks outperform vertex-only networks, giving a clue as to the area of difficulty).

{\bf Networks using track and vertex information outperform those with expert features.} Networks trained with track and vertex information but without the benefit of expert-level guidance and dimensional reduction manage to achieve better performance than those which use only expert-level features. This is remarkable, as the dimensionality of the tracks+vertices features is very large and expert-only networks represent the current state-of-the-art. Note, however, that for high signal efficiency ($>75$\%) the expert-only networks outperform the networks using tracks+vertices.

{\bf Networks which combine expert features with low-level information have the best performance.} Combining the lowest-level information for completeness with the low-dimensional hints from expert features significantly outperforms the state-of-the-art networks which use only expert features.  While in principle all of the information exists in the lowest-level features and it should be possible to train a network which matches or exceeds this performance without expert knowledge, this is neither necessary nor desirable.  Expert knowledge exists and is well-established, and there is no reason to discard it. 

In addition, this expert guidance encourages the network to identify discrimination strategies based on well-understood properties of the jet flavor problem and decreases the likelihood of relying on learning strategies based on spurious or poorly-modeled corners of the space.  We note that the use of high-dimensional lower-level data will require careful validation of the simulation models; reasonable strategies exist, such as a combination of the validation of individual features in one-dimensional projections with validation of the network output in control samples, which probes the use of information in multi-feature correlations.

These improvements in the performance of the tagger can give important boosts to physics studies which rely on the identification of jet flavor.


\section{Acknowledgements}

We thank David Kirkby, Gordon Watts, Shimon Whiteson, David Casper, and Kyle Cranmer  for useful comments and helpful discussion.   We thank Yuzo Kanomata for computing support. We also wish to acknowledge a hardware grant from NVIDIA and NSF grant  IIS-1321053 to PB.

\bibliographystyle{unsrt}
\bibliography{sadowski,physics}

\clearpage
\appendix

\end{document}